\documentclass[10pt,journal]{IEEEtran}
\usepackage{subcaption} 
\usepackage[final]{graphicx}
\usepackage{epsfig}
\usepackage{caption}
\usepackage{amsmath}
\usepackage{amssymb,times}
\usepackage{bm}
\usepackage{color}
\usepackage{cite}
\usepackage[ruled,vlined]{algorithm2e}     
\usepackage{epstopdf}
\usepackage[final]{graphicx}
\usepackage{epsfig}
\usepackage{subcaption}
\usepackage{seqsplit}
\usepackage{rotating}
\usepackage{booktabs}
\usepackage{tabularx} 
\usepackage{makecell} 
\usepackage{amsmath}
\usepackage{mathtools}
\usepackage{microtype}  
\usepackage{multirow}
\usepackage{tabularx}
\usepackage{booktabs}
\usepackage{makecell}
\usepackage{amsmath}
\usepackage{mathtools}

\IEEEoverridecommandlockouts

\begin{document}

\title{Adaptive Subspace Signal Detection and Performance Analysis in Nonzero-Mean Clutter}
\author{Weijian Liu, \emph{Senior Member, IEEE}, Zhenyu Xu, Jun Liu, \emph{Senior Member, IEEE}, Hui Chen, and \\
Yongxiang Liu, \emph{Member, IEEE}

\thanks{This work was supported by National Natural Science Foundation of China (Grant Nos. 62471485 and 62471450), Natural Science Foundation of Hubei Province (Grant No. 2025AFB873).}

\thanks{W. Liu, Z. Xu, and H. Chen are with Wuhan Electronic Information Institute, Wuhan 430019, China (e-mails: liuvjian@163.com, 1935620989@qq.com, and chhglr@sina.com).}
\thanks{J. Liu is with Department of Electronic Engineering and Information Science, University of Science and Technology of China, Hefei,  {\rm 230027}, China (e-mail: junliu@ustc.edu.cn).}
\thanks{Y. Liu is with the College of Electronic Science and Technology, National University of Defense Technology, Changsha 410073, China (e-mail: lyx\_bible@sina.com).}
\thanks{Digital Object Identifier 10.1109/TSP.2026.3692130}

}

\maketitle

\begin{abstract}
To solve the problem of detecting subspace signals in nonzero-mean clutter, we propose adaptive detectors, based on the strategies of generalized likelihood ratio test (GLRT), Rao test, Wald test, gradient test, and Durbin test. The results show that the detectors based on GLRT, Rao and Wald are structurally consistent with the subspace detectors in zero-means clutter. The analytic expressions for the probability of detection (PD) and probability of false alarm (PFA) of each detector are derived, and two major performance differences in the nonzero-mean clutter scenario are revealed. One is the loss of degree of freedom (DOF), which is reduced by 1 compared with the zero-mean clutter scenario. The second is the loss of signal-to-clutter (SCR) ratio. Simulation and measured data verify the effectiveness of the proposed detectors and demonstrate their practical value in real-world radar systems. 
\end{abstract}

\section{introduction}
Multichannel adaptive signal detection plays an irreplaceable role in radar, sonar, wireless communication and other fields \cite{LiuLiu22SCIS}. 
Since Kelly proposed the first adaptive detector based on the generalized likelihood ratio test (GLRT) in 1986, denoted as Kelly's GLRT (KGLRT)\cite{Kelly86}, adaptive detection under zero-mean clutter has formed a relatively complete theoretical system \cite{DeMaioGreco16Book,LiuOrlando22Book,Hao22Book,WangLiu24Book}. In particular, based on the KGLRT, the adaptive matched filter (AMF) \cite{ChenReed91}, and De Maio's Rao (DMRao) \cite{DeMaio07}, adaptive coherence estimator (ACE) \cite{KrautScharf99} are constantly proposed. Nowadays, adaptive detection technology has been used in advanced radars such as frequency diverse array (FDA) radar \cite{MiaoZhang24,Li2025,Lan2025} and multiple-input multiple-output (MIMO) radar \cite{Wang2024,Zhou2025}.

The aforementioned detectors are primarily for rank-one signals, which are characterized by well-defined steering vectors. In practice, however, signals may occupy a given subspace rather than a single direction. Such signals are referred to as subspace signals. Notably, a rank-one signal represents a special case of a subspace signal when the subspace dimension equals one.
The subspace signal model offers broader applicability. For instance, echoes from helicopter targets \cite{GiniFarina99} and polarimetric radar returns \cite{LiuGuo25DSP} can both be effectively modeled as subspace signals. Subspace generalizations of the KGLRT, AMF, DMRao, and ACE detectors have been developed in \cite{RaghavanPulsone96}, \cite{LiuZhang12b}, \cite{LiuXie14b}, and \cite{KrautScharf01}, respectively. For brevity, the subspace-based variants of the first three detectors are termed the subspace GLRT (SGLRT), subspace AMF (SAMF), and subspace Rao (SRao).

However, the aforementioned findings primarily focus on zero-mean clutter, while real-world scenarios may involve nonzero-mean clutter. For instance, 
in hyperspectral imaging, the mean spectral radiation of background targets is modulated by illumination conditions, atmospheric transmission and other factors, showing nonzero mean characteristics \cite{Vincent2020}. 

Moreover, in space-time adaptive processing (STAP) or array signal processing, the presence of outliers or interfering targets can cause the data to have a nonzero mean without the desired target being present. Outliers are typically caused by discrete clutter or interference \cite{Bassak2023}, while interfering targets originate from the sidelobe of radar receivers \cite{HanDeMaio19AES}.
In these scenarios, ignoring the nonzero mean of clutter can lead to the loss of the constant false alarm rate (CFAR) feature of the detector, or a severe degradation of probability of detection (PD).

The detection problem in nonzero-mean clutter modeled as elliptically symmetric distributions is investigated in \cite{FronteraPons2017}. Robust estimates of the clutter's covariance matrix and mean vector are obtained using fixed point estimators (FPE), and founded on these estimates, an effective detector is developed utilizing the adaptive normalized matched filter (ANMF) structure.

Furthermore, the same study derives the probability of false alarm (PFA) expression for this detector. For the signal detection problem in nonzero-mean Gaussian clutter, an adaptive detector with joint estimation of the mean and covariance matrix is constructed, and the analytical PFA is derived in \cite{FronteraPonsPascal17}.

In addressing the hyperspectral signal detection challenge in nonzero-mean clutter, a detector based on the ACE structure is designed in \cite{Vincent2020} given that both the nonzero-mean vector and covariance matrix of the clutter are known. A matrix-variate t-distribution is employed in \cite{OlivierFrancois20SP} to model the nonzero-mean non-Gaussian clutter, where a GLRT is derived. The results indicate that this detector is an extension of the GLRT for nonzero-mean Gaussian clutter.

The hyperspectral target detection problem in nonzero-mean matrix-variate t-distribution clutter is investigated in \cite{OlivierFrancois21SP}. By utilizing two sets of training data with different means but the same covariance matrix, an effective GLRT-based detector is developed. For the hyperspectral signal detection problem, the statistical performance of the real-valued KGLRT, ACE, and AMF in nonzero-mean clutter is analyzed in \cite{Besson2022} under two mismatch scenarios: an additive model and a replacement model. The former refers to the case where a mismatch is observed between the clutter means of the training and test data, while the latter involves an unknown amplitude mismatch in the clutter mean and an unknown mismatch in the covariance matrix between the training and test data.

To address the detection challenge in more complex environments, the detection problem in nonzero-mean compound-Gaussian sea clutter is studied in \cite{Wu2025}. Based on the GLRT, detectors are proposed for cases where the texture component follows inverse Gaussian, Gamma, and inverse Gamma distributions. The corresponding Rao and Wald detectors are provided in \cite{Wu2025a}.

It should be noted that although numerous studies have addressed target detection in nonzero-mean clutter, none have yet investigated subspace signal detection in such environments. For signal detection in nonzero-mean clutter, due to the excessive number of unknowns, an optimal detector does not exist. Beyond the GLRT criterion, several other effective detector design strategies are available, including the Wald, Rao, Durbin, and Gradient tests. In some scenarios, these strategies can even outperforms the GLRT in detection performance \cite{Xue22GRS,Besson23TSP,Besson25SPL,Besson25SP}.
Accordingly, this paper extends the rank-one signal model in \cite{FronteraPonsPascal17} to a subspace signal model in the complex domain, designing detectors based on multiple strategies and evaluating their statistical performance.

The most closely related prior studies are \cite{FronteraPonsPascal17} and \cite{Besson2022}, yet our work is distinguished from them in several key aspects. First, although effective detectors for nonzero-mean clutter have been proposed in \cite{FronteraPonsPascal17}, their design is primarily based on the GLRT, and the detectors derived are not strictly GLRTs in the true sense. Furthermore, with regard to statistical performance analysis, only the PFA expressions have been derived in \cite{FronteraPonsPascal17}, with no provision of the PDs.
When compared with the work carried out by Professors Besson and Vincent as presented in \cite{Besson2022}, significant differences are observed in this paper in the following key aspects. 1) The signal model employed in \cite{Besson2022} assumes a rank-one structure, whereas a subspace signal model, which encompasses the rank-one signal model as a special case, is adopted in our study. 2) The primary emphasis is placed on analyzing the statistical performance of existing detectors in \cite{Besson2022}, while new detectors are proposed and a comprehensive statistical performance analysis is conducted in this paper. 3) The proposed detectors and their statistical analysis are developed in the complex domain in this paper, while operations are carried out in the real domain in \cite{Besson2022}. 4) Detection performance under clutter mean mismatch or covariance matrix power mismatch is evaluated in \cite{Besson2022}, whereas detection performance under subspace signal mismatch is examined in this paper.

Clutter mean mismatch refers to inconsistency between the clutter mean in test and training data. Covariance matrix power mismatch indicates identical covariance structures but different power levels between test and training data. Subspace signal mismatch occurs when the signal component in test data does not fully reside within the assumed signal subspace. Practical factors like array errors, inhomogeneous propagation media, and target maneuvers may cause such signal mismatch \cite{LiuLiu20a,Huang24DSP}.

We have made three innovative contributions. 1) We develop effective adaptive detectors based on multiple detection strategies including the GLRT, Rao test, Wald test, gradient test, and Durbin test. Our analysis reveals important structural relationships. The GLRT-based, Rao-based, and Wald-based detectors are structurally consistent\footnote{``Structurally consistent'' means that the final algebraic form of the test statistic (e.g., the proposed detector in \eqref{eq55}) is the same as that of its zero-mean counterpart in \eqref{eq105}, with the only difference being that the test data vector $\textbf{x}$ and the sample covariance estimate $\textbf{S}$ are replaced by their mean-removed versions $\mathbf{z}$ defined in \eqref{eq54} and $\textbf{S}_2$ defined in \eqref{eq22}, respectively. This is important because it shows a deep connection between the problems and allows for the reuse of known statistical properties with minor modifications.}
 with their zero-mean counterparts (namely the SGLRT, SRao, and SAMF), while the gradient-based test and GLRT yield identical test statistics, and the Durbin-based test statistics shows equivalence to the Rao test statistics.
2) We obtain the statistical characterizations of these detectors, deriving analytical expressions for both PDs and PFAs.
3) We carry out comparative performance analysis identifying two fundamental differences from conventional zero-mean scenarios: (i) a degree of freedom (DOF) reduction in the test statistic's distribution (decreased by 1) due to joint mean and covariance matrix estimation, and (ii) a signal-to-clutter ratio (SCR) degradation with scaling factor $L/(L+1)$ (where $L$ represents training data size) caused by the estimation bias in nonzero-mean conditions.

This paper is organized in the following structure. The mathematical framework for the detection problem is rigorously developed in Section II. Section III presents the detailed derivations of the proposed detectors. Section IV conducts statistical performance analysis of the detectors, which includes: first, providing equivalent forms of three detectors\footnote{Two detectors are considered equivalent when they achieve identical PD at the same PFA. Generally speaking, equivalent detectors are monotonic increasing functions of each other \cite{LiuLiu22SCIS}.}, and subsequently deriving analytical expressions for PDs and PFAs. In Section V, we conduct a comprehensive performance evaluation of the proposed detectors using Monte Carlo simulations and three real datasets. Finally, Section VI summarizes the key findings and contributions of this paper.

\section{Detection Problem Description}

The test data $\mathbf{x}$ is an $N\times 1$ complex random vector, which is assumed to be distributed under one of two hypotheses. Under ${{\text{H}}_{0}}$, $\mathbf{x}$ contains only a non-zero-mean complex Gaussian clutter vector, denoted as   $\mathbf{c}\sim\mathcal{C}{{\mathcal{N}}_{N}}(\bm{\mu},\mathbf{R})$, where $\bm{\mu}$ is non-zero mean and $\mathbf{R}$ is the covariance matrix. Under ${{\text{H}}_{1}}$, $\mathbf{x}$ contains both the clutter component $\mathbf{c}$ and a signal component $\mathbf{A\bm{\alpha}}$, where $\mathbf{A}$ is an $N\times p$ full-rank signal matrix and $\mathbf{\bm{\alpha}}$ is a $p\times 1$ signal coordinate vector, and $p< N$.
In a radar context, $\mathbf{A}$ could represent the known Doppler or angle steering vectors of interest, while $\bm\alpha$ captures the unknown complex amplitudes, e.g., target reflectivity.
In addition to the test data, we assume the availability of $L$ independent and identically distributed (IID) training data ${{\mathbf{x}}_{\ell}}$, ${\ell}=1,2,\cdots,L$, where each ${{\mathbf{x}}_{\ell}}$ contains only the clutter component ${{\mathbf{c}}_{\ell}}$ sharing the same statistical properties as $\mathbf{c}$, i.e., ${{\mathbf{c}}_{\ell}}\sim\mathcal{C}{{\mathcal{N}}_{N}}(\bm{\mu},\mathbf{R})$.
Based on the above analysis, the detection task admits a binary hypothesis testing formulation
\begin{equation}
\label{eq1}
\begin{cases}\mathrm{H}_0:\mathbf{x}=\mathbf{c},\mathbf{x}_{\ell}=\mathbf{c}_{\ell},{\ell}=1,2,\cdots,L\\ \mathrm{H}_1:\mathbf{x}=\mathbf{A}\bm{\alpha}+\mathbf{c},\mathbf{x}_{\ell}=\mathbf{c}_{\ell},{\ell}=1,2,\cdots,L. \end{cases}
\end{equation}

In \eqref{eq1}, the unknown parameters implicitly include $\bm{\mu}$, $\mathbf{R}$, and $\mathbf{\bm{\alpha}}$. Note that the training data $\mathbf{x}_{\ell}=\mathbf{c}_{\ell},{\ell}=1,2,\cdots,L$ are identical under both hypotheses and represent signal-absent observations. Due to the presence of these unknown quantities, an optimal detector does not exist. Therefore, it is imperative to develop detectors utilizing multiple strategies and subsequently compare their performance.

\section{Detector Design}

The GLRT, Wald, and Rao tests are three essential strategies for designing detectors. In recent years, the Durbin and Gradient tests have also been widely employed in detector design \cite{SunLiu22TSP}. 
In the subsequent content, we successively derive the detectors based on GLRT, Rao test, and Wald test\footnote{
The detector derived from the gradient test is equivalent to its GLRT-based counterpart (shown in Appendix \ref{app:GradTest}), while the detector based on the Durbin test is equivalent to the one based on the Rao test (shown in Appendix \ref{app:DurbinTest}).}.

\subsection{GLRT-Based Detector}

The GLRT can be expressed as
\begin{equation}
	\label{eq2}
	{{t}_{\text{GLRT}}}=\frac{\underset{\mathbf{R},\bm{\mu} ,\mathbf{\bm{\alpha} }}{\mathop{\max }}\,{{f}_{1}}(\mathbf{x},{{\mathbf{X}}_{L}})}{\underset{\mathbf{R},\bm{\mu} }{\mathop{\max }}\,{{f}_{0}}(\mathbf{x},{{\mathbf{X}}_{L}})},
\end{equation}
where ${\mathbf{X}}_{L}=[{\mathbf{x}}_{1},{\mathbf{x}}_{2},\cdots,{\mathbf{x}}_{L}]$, $ {{f}_{0}}(\mathbf{x},{{\mathbf{X}}_{L}}) $ and $ {{f}_{1}}(\mathbf{x},{{\mathbf{X}}_{L}}) $ represent the joint probability density functions (PDFs) of the test and training data under hypotheses ${{\text{H}}_{0}}$ and ${{\text{H}}_{1}}$, respectively, given by
\begin{equation}
	\label{eq3}
	\begin{aligned}
	{{f}_{0}}(\mathbf{x},{{\mathbf{X}}_{L}})&=
{\left[{{\pi }^{N(L+1)}}|\mathbf{R}{{|}^{L+1}}\right]}^{-1}\\
&\cdot{\text{e}^{{- \big[{{( \mathbf{x}-\bm{\mu} )}^{\text{H}}}{{\mathbf{R}}^{-1}}( \mathbf{x}-\bm{\mu} )} {+\sum\limits_{{\ell}=1}^{L}{{{( {{\mathbf{x}}_{\ell}}-\bm{\mu}  )}^{\text{H}}}{{\mathbf{R}}^{-1}}( {{\mathbf{x}}_{\ell}}-\bm{\mu}  )} \big]}}},
\end{aligned}
\end{equation}
and
\begin{equation}
	\label{eq4}
	\begin{aligned}
	{{f}_{1}}(\mathbf{x},{{\mathbf{X}}_{L}})=
&{\left[{{\pi }^{N(L+1)}}|\mathbf{R}{{|}^{L+1}}\right]}^{-1}\\
&\cdot{\text{e}^{{- \big[{ {\mathbf{x}}_{\text{rs}}^{\text{H}}} {{\mathbf{R}}^{-1}} {{\mathbf{x}}_{\text{rs}}} } {+\sum\limits_{{\ell}=1}^{L}{{{( {{\mathbf{x}}_{\ell}}-\bm{\mu}  )}^{\text{H}}}{{\mathbf{R}}^{-1}}( {{\mathbf{x}}_{\ell}}-\bm{\mu}  )} \big]}}},
	\end{aligned}
\end{equation}
where
\begin{equation}
\label{x1}
{{\mathbf{x}}_{\text{rs}}}=\mathbf{x}-\mathbf{A\bm{\alpha} }-\bm{\mu}.
\end{equation}

It is shown in Appendix \ref{app:DrvtGLRTNMC}
that the GLRT is found to be
\begin{equation}
	\label{eq27}
	\begin{aligned}
	&{{t}_{\text{GLRT}}}=\frac{\left| {{\mathbf{S}}_{0}} \right|}{\left| {{\mathbf{S}}_{2}} \right|}\left[ 1+{{\left( \mathbf{x}-{{{\hat{\bm{\mu} }}}_{0}} \right)}^{\text{H}}}\mathbf{S}_{0}^{-1}\left( \mathbf{x}-{{{\hat{\bm{\mu} }}}_{0}} \right) \right]\\&\cdot\left\{ 1+\frac{L+1}{L}\left[{{\left(\mathbf{x}-{{{\hat{\bm{\mu} }}}_{0}} \right)}^{\text{H}}}\mathbf{S}_{2}^{-1}\left( \mathbf{x}-{{{\hat{\bm{\mu} }}}_{0}} \right)\right.\right.\\&\left.\left.-{{\left( \mathbf{x}-{{{\hat{\bm{\mu} }}}_{0}} \right)}^{\text{H}}}\mathbf{S}_{2}^{-1}\mathbf{A}{{({{\mathbf{A}}^{\text{H}}}\mathbf{S}_{2}^{-1}\mathbf{A})}^{-1}}{{\mathbf{A}}^{\text{H}}}\mathbf{S}_{2}^{-1}\left( \mathbf{x}-{{{\hat{\bm{\mu} }}}_{0}} \right) \right] \right\}^{-1},
	\end{aligned}
\end{equation}
where\footnote{As will be shown later, the detectors based on both the GLRT and Wald test depend on the positive definiteness of $\mathbf{S}_2$. Appendix \ref{app:S2} proves that when $L \geq N+1$, $\mathbf{S}_2$ is positive definite with probability 1.}
\begin{equation}
	\label{eq9}
	{{\mathbf{S}}_{0}}=\sum\limits_{{\ell}=1}^{L}{\left( {{\mathbf{x}}_{\ell}}-{{{\hat{\bm{\mu} }}}_{0}} \right){{\left( {{\mathbf{x}}_{\ell}}-{{{\hat{\bm{\mu} }}}_{0}} \right)}^{\text{H}}}},
\end{equation}
\begin{equation}
	\label{eq22}
	{{\mathbf{S}}_{2}}={{\mathbf{S}}_{0}}-\frac{1}{L}\left( \mathbf{x}-{{{\hat{\bm{\mu} }}}_{0}} \right){{\left( \mathbf{x}-{{{\hat{\bm{\mu} }}}_{0}} \right)}^{\text{H}}},
\end{equation}
and
\begin{equation}
	\label{eq7}
	{{\hat{\bm{\mu} }}_{0}}=\frac{1}{L+1}\left( \mathbf{x}+\sum\limits_{{\ell}=1}^{L}{{{\mathbf{x}}_{\ell}}} \right).
\end{equation}
Note that ${{\mathbf{S}}_{0}}$ can be taken as the sample covariance matrix (SCM) with the nonzero mean ${\bm{\mu} }$ being removed under hypothesis $\text{H}_0$.

For notational convenience, we refer to the detector in \eqref{eq27} as the subspace-based GLRT for nonzero mean clutter (SGLRT-NMC).

It is shown in Appendix \ref{app:GLRTNMC} that \eqref{eq27} is equivalent to
\begin{equation}
	\label{eq55}
	\begin{aligned}
	&t_{\text{SGLRT-NMC}}^{'}\\&=\frac{{{\mathbf{z}}^{\text{H}}}\mathbf{S}_{2}^{-1}\mathbf{A} {{({{\mathbf{A}}^{\text{H}}}\mathbf{S}_{2}^{-1}\mathbf{A})}^{-1}}{{\mathbf{A}}^{\text{H}}} \mathbf{S}_{2}^{-1}\mathbf{z}}{1+{{\mathbf{z}}^{\text{H}}}\mathbf{S}_{2}^{-1}\mathbf{z}- {{\mathbf{z}}^{\text{H}}}\mathbf{S}_{2}^{-1}\mathbf{A}{{({{\mathbf{A}}^{\text{H}}}\mathbf{S}_{2}^{-1} \mathbf{A})}^{-1}}{{\mathbf{A}}^{\text{H}}}\mathbf{S}_{2}^{-1}\mathbf{z}},
	\end{aligned}
\end{equation}
where
\begin{equation}
	\label{eq54}
	\mathbf{z}=\sqrt{\frac{L+1}{L}}\left( \mathbf{x}-{{{\hat{\bm{\mu} }}}_{0}} \right).
\end{equation}
Compared with \eqref{eq27}, \eqref{eq55} is more convenient for deriving the statistical properties of the SGLRT-NMC. The specific analysis  will be elaborated in detail in next section.

\subsection{Rao Test-Based Detector}
The Rao test is formulated as
\begin{equation}
	\label{eq28}
	{{t}_{\text{Rao}}}\!=\!\!\!{\left. \left. \frac{\partial \ln {{f}_{1}}(\mathbf{x},\mathbf{X}_L)} {\partial {{\mathbf{\Theta }}_{\text{r}}}} \right|_{\mathbf{\Theta }={{{\mathbf{\hat{\Theta }}}}_{0}}}^{\text{T}}\!{\!{\left[ {{\mathbf{I}}^{-1}}( {{{\mathbf{\hat{\Theta }}}}_{0}} ) \right]}_{{{\mathbf{\Theta }}_{\text{r}}},{{\mathbf{\Theta }}_{\text{r}}}}}\!\!\frac{\partial \ln {{f}_{1}}(\mathbf{x},\mathbf{X}_L)}{\partial \mathbf{\Theta }_{\text{r}}^{*}} \right|}_{\mathbf{\Theta }={{{\mathbf{\hat{\Theta }}}}_{0}}},
\end{equation}
where $ {{\left[ {{\mathbf{I}}^{-1}}({{{\mathbf{\hat{\Theta }}}}_{0}}) \right]}_{{{\mathbf{\Theta }}_{\text{r}}},{{\mathbf{\Theta }}_{\text{r}}}}}$ represents the $\left( {{\mathbf{\Theta }}_{\text{r}}},{{\mathbf{\Theta }}_{\text{r}}} \right)$ block of the inverse Fisher information matrix (FIM) $\mathbf{I}({{\mathbf{\hat{\Theta }}}_{0}})$, and
\begin{equation}
	\label{eq29}
	\mathbf{I}(\mathbf{\Theta })=\text{E}\left[ \frac{\partial \ln f(\mathbf{x};\mathbf{\Theta })}{\partial {{\mathbf{\Theta }}^{*}}}\frac{\partial \ln f(\mathbf{x};\mathbf{\Theta })}{\partial {{\mathbf{\Theta }}^{\text{T}}}} \right].
\end{equation}

For the detection problem in \eqref{eq1}, let
\begin{equation}
	\label{eq30}
	\mathbf{\Theta }={{\left[ \mathbf{\Theta }_{\text{r}}^{\text{T}},\mathbf{\Theta }_{\text{s}}^{\text{T}} \right]}^{\text{T}}},
\end{equation}
is $\left( {{N}^{2}}+N+1 \right)\times 1$ parameter vector, where ${{\mathbf{\Theta }}_{\text{r}}}=\mathbf{\bm{\alpha} }$ and ${{\mathbf{\Theta }}_{\text{s}}}={{\left[ {{\bm{\mu} }^{\text{T}}},{{\operatorname{vec}}^{\text{T}}}(\mathbf{R}) \right]}^{\text{T}}}$.

It is shown in Appendix \ref{app:DrvtRaoNMC} the Rao test is given by
\begin{equation}
	\label{eq45}
	\begin{aligned}t_{\mathrm{Rao}}&= \left(\mathbf{x}-\hat{\mathbf{\bm{\mu}}}_0\right)^\mathrm{H} \left[\left(\mathbf{x}-\hat{\mathbf{\bm{\mu}}}_0\right) \left(\mathbf{x}-\hat{\mathbf{\bm{\mu}}}_0\right)^\mathrm{H}+\mathbf{S}_0\right]^{-1}\\ &\cdot\mathbf{A}\left\{\mathbf{A}^\mathrm{H}\left[\left(\mathbf{x}- \hat{\mathbf{\bm{\mu}}}_0\right)\left(\mathbf{x}-\hat{\mathbf{\bm{\mu}}}_0\right)^\mathrm{H} +\mathbf{S}_0\right]^{-1}\mathbf{A}\right\}^{-1}\\&\cdot\mathbf{A}^{\mathrm{H}} \left[\left(\mathbf{x}-\hat{\mathbf{\bm{\mu}}}_0\right)\left(\mathbf{x}- \hat{\mathbf{\bm{\mu}}}_0\right)^{\mathrm{H}}+\mathbf{S}_0\right]^{-1} \left(\mathbf{x}-\hat{\mathbf{\bm{\mu}}}_0\right).\end{aligned}
\end{equation}

For notational convenience, we refer to the detector in \eqref{eq45} as the subspace-based Rao test statistic for nonzero mean clutter (SRao-NMC).

It is shown in Appendix \ref{app:RaoNMC} that \eqref{eq45} can be expressed as
\begin{equation}
	\label{eq60}
	\begin{aligned}
	&t_{\mathrm{SRao-NMC}}= {(1+\mathbf{z}^{\mathrm{H}}\mathbf{S}_{2}^{-1}\mathbf{z})}^{-1}\\ &\cdot\frac{\mathbf{z}^{\mathrm{H}}\mathbf{S}_{2}^{-1}\mathbf{A} (\mathbf{A}^{\mathrm{H}}\mathbf{S}_{2}^{-1}\mathbf{A})^{-1}\mathbf{A}^{\mathrm{H}} \mathbf{S}_{2}^{-1}\mathbf{z}} {1+\mathbf{z}^{\mathrm{H}}\mathbf{S}_{2}^{-1}\mathbf{z} -\mathbf{z}^{\mathrm{H}}\mathbf{S}_{2}^{-1}\mathbf{A}(\mathbf{A}^{\mathrm{H}}\mathbf{S}_{2}^{-1} \mathbf{A})^{-1}\mathbf{A}^{\mathrm{H}}\mathbf{S}_{2}^{-1}\mathbf{z}} .
	\end{aligned}
\end{equation}
\subsection{Wald Test-Based Detector}

The Wald test can be expressed as
\begin{equation}
	\label{eq46}
	{{t}_{\text{Wald}}}={{\left( {{{\mathbf{\hat{\Theta }}}}_{{{\text{r}}_{1}}}}-{{\mathbf{\Theta }}_{{{\text{r}}_{0}}}} \right)}^{\text{H}}}{{\left\{ {{\left[ {{\mathbf{I}}^{-1}}\left( {{{\mathbf{\hat{\Theta }}}}_{1}} \right) \right]}_{{{\mathbf{\Theta }}_{\text{r}}},{{\mathbf{\Theta }}_{\text{r}}}}} \right\}}^{-1}}\left( {{{\mathbf{\hat{\Theta }}}}_{{{\text{r}}_{1}}}}-{{\mathbf{\Theta }}_{{{\text{r}}_{0}}}} \right).
\end{equation}
%

Substituting \eqref{eq25} and \eqref{eq42} into \eqref{eq46} and dropping the constant yields the Wald test for known $\mathbf{R}$ as
\begin{equation}
	\label{eq48}
	\begin{aligned}
	{{t}_{\text{Wal}{{\text{d}}_{\mathbf{R}}}}}&={{\left( \mathbf{x}-{{{\hat{\bm{\mu} }}}_{0}} \right)}^{\text{H}}}\mathbf{S}_{2}^{-1}\mathbf{A}({{\mathbf{A}}^{\text{H}}}\mathbf{S}_{2}^{-1}\mathbf{A})_{{}}^{-1}\\&\cdot{{\mathbf{A}}^{\text{H}}}\mathbf{R}_{{}}^{-1}\mathbf{A}({{\mathbf{A}}^{\text{H}}}\mathbf{S}_{2}^{-1}\mathbf{A})_{{}}^{-1}{{\mathbf{A}}^{\text{H}}}\mathbf{S}_{2}^{-1}\left( \mathbf{x}-{{{\hat{\bm{\mu} }}}_{0}} \right).
	\end{aligned}
\end{equation}
By replacing $\mathbf{R}$ in \eqref{eq48} with its MLE under ${{\text{H}}_{1}}$ and omitting constant terms, we obtain the Wald test statistic as
\begin{equation}
	\label{eq49}
	\begin{aligned}
	{{t}_{\text{Wald}}}&={{\left( \mathbf{x}-{{{\hat{\bm{\mu} }}}_{0}} \right)}^{\text{H}}}\mathbf{S}_{2}^{-1}\mathbf{A}({{\mathbf{A}}^{\text{H}}}\mathbf{S}_{2}^{-1}\mathbf{A})_{{}}^{-1}\\&\cdot{{\mathbf{A}}^{\text{H}}}\mathbf{\hat{R}}_{1}^{-1}\mathbf{A}({{\mathbf{A}}^{\text{H}}}\mathbf{S}_{2}^{-1}\mathbf{A})_{{}}^{-1}{{\mathbf{A}}^{\text{H}}}\mathbf{S}_{2}^{-1}\left( \mathbf{x}-{{{\hat{\bm{\mu} }}}_{0}} \right).
	\end{aligned}
\end{equation}

For notational simplicity, we refer to the detector in \eqref{eq49} as the subspace-based adaptive matched filter for nonzero mean clutter (SAMF-NMC).

It is shown in Appendix \ref{app:WaldNMC} that \eqref{eq49} can be rewritten as
\begin{equation}
	\label{eq65}
	{{t}_{\text{SAMF-NMC}}}={{\mathbf{z}}^{\text{H}}}\mathbf{S}_{2}^{-1}\mathbf{A} ({{\mathbf{A}}^{\text{H}}} \mathbf{S}_{2}^{-1}\mathbf{A})_{{}}^{-1}{{\mathbf{A}}^{\text{H}}} \mathbf{S}_{2}^{-1}\mathbf{z}.
\end{equation}
\section{Statistical Performance Analysis}

During the statistical analysis phase, we consider a more generalized application scenario - signal mismatch conditions \cite{LiuLiu16SP}, where the true signal steering vector ${{\mathbf{p}}_{0}}$ is not entirely contained within the assumed signal subspace generated by the signal matrix $\mathbf{A}$. Signal matching can be viewed as a particular instance of signal mismatch with zero mismatch angle.

\subsection{The PD and PFA of the SGLRT-NMC in \eqref{eq55}}

The form of the SGLRT-NMC in \eqref{eq55} facilitates derivation of its statistical distribution. We now sequentially analyze the statistical characteristics of $\mathbf{z}$ and ${{\mathbf{S}}_{2}}$ in \eqref{eq55}. Substituting \eqref{eq7} into \eqref{eq54} yields
\begin{equation}
	\label{eq66}
	\begin{aligned}
	\mathbf{z}=\sqrt{\frac{L+1}{L}}\left[\frac{L}{L+1}\mathbf{x}-\frac{1}{L+1} \sum_{{\ell}=1}^L\mathbf{x}_{\ell}\right].
	\end{aligned}
\end{equation}
According to \eqref{eq66}, the mean values of $ \mathbf{z} $ under hypotheses ${{\text{H}}_{0}}$ and ${{\text{H}}_{1}}$ are respectively given by
\begin{equation}
	\label{eq67}
	\begin{aligned}\mathrm{E}\left(\mathbf{z};\mathrm{H}_{0}\right)=\mathbf{0}
\end{aligned}
\end{equation}
and
\begin{equation}
	\label{eq68}
	\begin{aligned}\mathrm{E}\left(\mathbf{z};\mathrm{H}_{1}\right) =\sqrt{\frac{L}{L+1}}\mathbf{A}\mathbf{\bm{\alpha}}.
\end{aligned}
\end{equation}
Moreover, as indicated by \eqref{eq66}, the covariance matrix of $ \mathbf{z} $ remains identical under both ${{\text{H}}_{0}}$ and ${{\text{H}}_{1}}$, specifically
\begin{equation}
	\label{eq69}
	\begin{aligned}\mathrm{Cov}(\mathbf{z})=\mathbf{R}, 
\end{aligned}
\end{equation}
where $\text{Cov}\left( \cdot \right)$ denotes the covariance operator. It follows from \eqref{eq67} and \eqref{eq69} that 
\begin{equation}
	\label{eq70}
	\mathbf{z}\mid\text{H}_0\sim\mathcal{CN}_N(\mathbf{0},\mathbf{R}),~ \mathbf{z}\mid\text{H}_1\sim\mathcal{CN}_N\left(\sqrt{\tfrac{L}{L+1}}\mathbf{A}\mathbf{\bm{\alpha}},\mathbf{R}\right).
\end{equation}
According to \cite{KellyForsythe89} and Theorem 3.1.2 in \cite{Muirhead05}, the matrix ${{\mathbf{S}}_{2}}$ shown in \eqref{eq122} follows a complex Wishart distribution with $L-1$ DOFs and covariance matrix $\mathbf{R}$ under both hypotheses ${{\text{H}}_{0}}$ and ${{\text{H}}_{1}}$, denoted as
\begin{equation}
	\label{eq71}
	{{\mathbf{S}}_{2}}\sim \mathcal{C}\mathcal{W}\left( L-1,\mathbf{R} \right).
\end{equation}
A loss factor can be defined as
\begin{equation}
	\label{eq72}
	\beta ={{\left[ 1+{{\mathbf{z}}^{\text{H}}}\mathbf{S}_{2}^{-1}\mathbf{z}-{{\mathbf{z}}^{\text{H}}}\mathbf{S}_{2}^{-1}\mathbf{A}{{({{\mathbf{A}}^{\text{H}}}\mathbf{S}_{2}^{-1}\mathbf{A})}^{-1}}{{\mathbf{A}}^{\text{H}}}\mathbf{S}_{2}^{-1}\mathbf{z} \right]}^{-1}}.
\end{equation}

At this point, it can be observed that the SGLRT-NMC in \eqref{eq55} and the loss factor in \eqref{eq72} share identical structures with their counterparts in \cite{LiuLiu16SP} for the case of zero-mean clutter. Consequently, drawing on the findings established in \cite{LiuLiu16SP,KellyForsythe89}, the statistical performance of the SGLRT-NMC in \eqref{eq55} can be readily derived.
When $ \beta $ is given, a noncentral complex F-distribution with DOFs $p$ and $L-N$, and noncentrality parameter $\beta {{\rho }_{\theta }}$ describes the conditional statistical behavior of the SGLRT-NMC in \eqref{eq55} under hypothesis ${{\text{H}}_{1}}$, denoted as
\begin{equation}
	\label{eq73}
	t_{\text{SGLRT-NMC}}^{'}\left| \left[ \beta ;{{\text{H}}_{1}} \right] \right.\sim \mathcal{C}{{\mathcal{F}}_{p,L-N}}\left( \beta {{\rho }_{\theta }} \right),
\end{equation}
where $ {{\rho }_{\theta }}=\rho {{\cos }^{2}}\theta  $,
\begin{equation}
	\label{eq74}
	{{\cos }^{2}}\theta =\frac{\mathbf{p}_{0}^{\text{H}}{{\mathbf{R}}^{-1}}\mathbf{A}{{({{\mathbf{A}}^{\text{H}}}{{\mathbf{R}}^{-1}}\mathbf{A})}^{-1}}{{\mathbf{A}}^{\text{H}}}{{\mathbf{R}}^{-1}}{{\mathbf{p}}_{0}}}{\mathbf{p}_{0}^{\text{H}}{{\mathbf{R}}^{-1}}{{\mathbf{p}}_{0}}},
\end{equation}
is the squared cosine of the angle between the true signal ${{\mathbf{p}}_{0}}$ and the signal subspace spanned by the matrix $\mathbf{A}$, and
\begin{equation}
	\label{eq75}
	\rho =\frac{L}{L+1}\mathbf{p}_{0}^{\text{H}}{{\mathbf{R}}^{-1}}{{\mathbf{p}}_{0}},
\end{equation}
serves as the noncentrality parameter, defined as the signal-to-clutter ratio (SCR).

Under the hypothesis ${{\text{H}}_{0}}$ where $\mathbf{\bm{\alpha}} = \mathbf{0}_{p\times 1}$, the statistical distribution of the SGLRT-NMC in \eqref{eq55} degenerates to
\begin{equation}
	\label{eq76}
	t_{\text{SGLRT-NMC}}^{'}\left| {{\text{H}}_{0}} \right.\sim \mathcal{C}{{\mathcal{F}}_{p,L-N}}.
\end{equation}
Based on \cite{LiuLiu16SP}, the loss factor $\beta$ shown in \eqref{eq72} follows a noncentral complex Beta distribution with DOFs $L-N+p$ and $N-p$, and noncentrality parameter ${{\delta}^{2}}$ under hypothesis ${{\text{H}}_{1}}$, denoted as
\begin{equation}
	\label{eq77}
	\beta |{{\text{H}}_{1}}\sim\mathcal{C}{{\mathcal{B}}_{L-N+p,N-p}}({{\delta }^{2}}),
\end{equation}
where
\begin{equation}
	\label{eq78}
	{{\delta }^{2}}=\rho {{\sin }^{2}}\theta,
\end{equation}
$ {{\sin }^{2}}\theta =1-{{\cos }^{2}}\theta $, and $ {{\cos }^{2}}\theta $ is given in \eqref{eq74}. Under hypothesis ${{\text{H}}_{0}}$, the distribution in \eqref{eq77} degenerates to its central form
\begin{equation}
	\label{eq79}
	\beta |{{\text{H}}_{0}}\sim\mathcal{C}{{\mathcal{B}}_{L-N+p,N-p}}.
\end{equation}
The DOFs $L-N+p$ and $N-p$ are tied to the dimensions of the signal subspace and the noise-only subspace, respectively, while the noncentrality parameter $\delta^2$ quantifies the energy of the signal component that lies outside the assumed signal subspace, directly linking it to the severity of signal mismatch.

Based on the aforementioned statistical distribution results, the PD of the SGLRT-NMC in \eqref{eq55} can be expressed as
\begin{equation}
	\label{eq80}
	\begin{aligned}\mathrm{PD}_{\mathrm{SGLRT-NMC}}&=\mathrm{Pr}\left[t_{\mathrm{SGLRT-NMC}}^{^{\prime}} >\eta_{\mathrm{G}};\mathrm{H}_{1}\right]\\ 
&=1-\int_0^1\mathcal{P}_1\left(\eta_\mathrm{G}\right)f_1\left(\beta\right)\mathrm{d}\beta, \end{aligned}
\end{equation}
where ${{\eta}_{\text{G}}}$ represents the detection threshold of the SGLRT-NMC in \eqref{eq55}, and ${{f}_{1}}(\beta)$ represents the PDF of $\beta$ given by
\begin{equation}
	\label{eq81}
	\begin{aligned}
	f_{1}\left(\beta\right)&=(L-N+p)(L-1)!\mathrm{e}^{-\delta^{2}\beta}\beta^{L-N+p-1}\\ &\cdot\sum_{k=0}^{L-N+p}\frac{\delta^{2k}(1-\beta)^{N-p+k-1}}{k!(L-N+p-k)!(N-p+k-1)!},
	\end{aligned}
\end{equation}
${{\mathcal{P}}_{1}}\left( {{\eta }_{\text{G}}} \right)$ represents the cumulative distribution function (CDF) of $ t_{\text{SGLRT-NMC}}^{'} $ conditioned on $\beta$ under ${{\text{H}}_{1}}$, expressed as
\begin{equation}
	\label{eq82}
	\begin{aligned}\mathcal{P}_{1}(\eta_{\mathrm{G}})& =\mathrm{Pr}\left[t_{\mathrm{SGLRT-NMC}}^{^{\prime}}\leq\eta_{\mathrm{G}}|\beta;\mathrm{H}_{1}\right]\\ &=\sum_{k=0}^{L-N-1}\mathrm{C}_{L-N+p-1}^{k+p}\frac{\eta_{\mathrm{G}}^{k+p}\mathrm{IG}_{k+1} \left(\frac{\rho_{\theta}\beta}{1+\eta_{\mathrm{G}}}\right)}{\left(1+\eta_{\mathrm{G}}\right)^{L-N+p-1}}.
\end{aligned}
\end{equation}
In \eqref{eq82}, $\text{C}_{L-N+p-1}^{k+p}$ denotes the binomial coefficient, and $\text{IG}_{k+1}(a)$ is the inverse gamma function, i.e.,
\begin{equation}
	\label{eq83}
	\text{I}{{\text{G}}_{k+1}}\left( a \right)={{e}^{-a}}\sum\limits_{m=0}^{k}{\frac{{{a}^{m}}}{m!}}.
\end{equation}

The PFA of the SGLRT-NMC in \eqref{eq55} is expressed as
\begin{equation}
	\label{eq84}
	\begin{aligned}\mathrm{PFA}_{\mathrm{SGLRT-NMC}}&=\mathrm{Pr}\left[t_{\mathrm{SGLRT-NMC}}^{^{\prime}}> \eta_{\mathrm{G}};\mathrm{H}_{0}\right]\\&=\int_{0}^{1}\left[1-\mathcal{P}_{0}\left(\eta_{\mathrm{G}}\right)\right] f_{0}\left(\beta\right)\mathrm{d}\beta,\end{aligned}
\end{equation}
where ${{\mathcal{P}}_{0}}( {{\eta }_{\text{G}}} )$ denotes the cumulative distribution function (CDF) of $ t_{\text{SGLRT-NMC}}^{'} $ conditioned on $\beta$ under  ${{\text{H}}_{0}}$, i.e.,
\begin{equation}
	\label{eq85}
	\begin{aligned}\mathcal{P}_{0}(\eta_{\mathrm{G}}) &=\mathrm{Pr}\left[t_{\mathrm{SGLRT-NMC}}^{^{\prime}} \leq\eta_{\mathrm{G}}|\beta;\mathrm{H}_{0}\right]\\ &=\sum_{k=0}^{L-N-1}\mathrm{C}_{L-N+p-1}^{k+p} \frac{\eta_{\mathrm{G}}^{k+p}}{\left(1+\eta_{\mathrm{G}}\right)^{L-N+p-1}}.
\end{aligned}
\end{equation}
The term ${{f}_{0}}\left( \beta \right)$ represents the PDF of $\beta$ defined in \eqref{eq79}, with its explicit expression given by
\begin{equation}
	\label{eq86}
	{{f}_{0}}\left( \beta  \right)=\frac{{{\beta }^{L-N+p-1}}{{\left( 1-\beta  \right)}^{N-p-1}}}{\text{B}(L-N+p,N-p)}.
\end{equation}
For convenience, we define
\begin{equation}
	\label{eq87}
	\begin{aligned}g\left(\eta_{\mathrm{G}}\right)&=1-\mathcal{P}_{0} \left(\eta_{\mathrm{G}}\right)\\&=1-\sum_{k=0}^{L-N-1}\mathrm{C}_{L-N}^{k+1} \frac{\eta_{\mathrm{G}}^{k+1}}{\left(1+\eta_{\mathrm{G}}\right)^{L-N}}.\end{aligned}
\end{equation}
Let $t=k+p$. Then, \eqref{eq87} can be expressed as
\begin{equation}
	\label{eq88}
	\begin{aligned}g\left(\eta_{\mathrm{G}}\right)&= 1-\sum_{t=p}^{L-N+p-1}\mathrm{C}_{L-N+p-1}^{t}\frac{\eta_{\mathrm{G}}^{t}} {\left(1+\eta_{\mathrm{G}}\right)^{L-N+p-1}}\\
&=\sum_{t=0}^{p-1}\mathrm{C}_{L-N+p-1}^{t}\frac{\eta_{\mathrm{G}}^{t}} {\left(1+\eta_{\mathrm{G}}\right)^{L-N+p-1}},
\end{aligned}
\end{equation}
where we utilized the binomial identity ${{\left( b+c \right)}^{N}}=\sum\nolimits_{n=0}^{N}{\text{C}_{N}^{n}{{b}^{n}}{{c}^{N-n}}}$. Substituting \eqref{eq88} into \eqref{eq84} yields
\begin{equation}
	\label{eq89}
	\text{PF}{{\text{A}}_{\text{SGLRT-NMC}}}=\sum\limits_{t=0}^{p-1}{\text{C}_ {L-N+p-1}^{t}\frac{\eta _{\text{G}}^{t}}{{{\left( 1+{{\eta }_{\text{G}}} \right)}^{L-N+p-1}}}}.
\end{equation}

%
\subsection{The PD and PFA of the SRao-NMC in \eqref{eq60}}

The SRao-NMC in \eqref{eq60} can be expressed as
\begin{equation}
	\label{eq92}
	t_{\text{SRao-NMC}}^{'}=\frac{\beta t_{\text{SGLRT-NMC}}^{'}}{1+t_{\text{SGLRT-NMC}}^{'}},
\end{equation}
where $ t_{\text{SGLRT-NMC}}^{'} $ and $ \beta $ are given by \eqref{eq55} and \eqref{eq72}, respectively.
Based on \eqref{eq92}, the expressions for calculating the PD and PFA of the SRao-NMC can be conveniently derived. Specifically, its PD can be rewritten as
\begin{equation}
	\label{eq93}
	\begin{aligned}\mathrm{PD_{SRao-NMC}}&=\mathrm{Pr}\left[\frac{\beta t_{\mathrm{SGLRT-NMC}}^{^{\prime}}}{1+t_{\mathrm{SGLRT-NMC}}^{^{\prime}}}>\eta_{\mathrm{R}}; \mathrm{H}_{1}\right]\\
&=\int_{\eta_{\mathrm{R}}}^{1}\left[1-\mathcal{P}_{1}\left(\frac{\eta_{\mathrm{R}}} {\beta-\eta_{\mathrm{R}}}\right)\right]f_{1}\left(\beta\right)\mathrm{d}\beta,\end{aligned}
\end{equation}
where $ {{\eta }_{\text{R}}} $ is the detection threshold for the SRao-NMC in \eqref{eq60}. By applying the same processing method as in \eqref{eq93}, the PFA of the SRao-NMC can be expressed as
\begin{equation}
	\label{eq94}
	\begin{gathered}\mathrm{PFA}_{\mathrm{SRao-NMC}}=\int_{\eta_{\mathrm{Rao}}}^{1} \left[1-\mathcal{P}_{0}\left(\frac{\eta_{\mathrm{R}}}{\beta-\eta_{\mathrm{R}}}\right)\right] f_{0}\left(\beta\right)\mathrm{d}\beta\\=\int_{\eta_{\mathrm{R}}}^{1}g \left(\frac{\eta_{\mathrm{R}}}{\beta-\eta_{\mathrm{R}}}\right)f_{0}\left(\beta\right) \mathrm{d}\beta. \end{gathered}
\end{equation}

\subsection{The PD and PFA of the SAMF-NMC in \eqref{eq65}}

The SAMF-NMC in \eqref{eq65} can be expressed as
\begin{equation}
	\label{eq101}
	{{t}_{\text{SAMF-NMC}}}=\frac{t_{\text{SGLRT-NMC}}^{'}}{\beta },
\end{equation}
where $ t_{\text{SGLRT-NMC}}^{'} $ and $ \beta $ are shown in equations \eqref{eq55} and \eqref{eq72}, respectively.
According to  \eqref{eq101}, the PD of the SAMF-NMC can be expressed as
\begin{equation}
	\label{eq102}
	\begin{aligned}\mathrm{PD}_{\mathrm{SAMF-NMC}}&=\mathrm{Pr}\left[\frac{t_{\mathrm{SGLRT-NMC}} ^{^{\prime}}}{\beta}>\eta_{\mathrm{W}};\mathrm{H}_{1}\right]\\
&=1-\int_0^1\mathcal{P}_1\left(\beta\eta_\mathrm{w}\right)f_1(\beta) \mathrm{d}\beta,\end{aligned}
\end{equation}
where $ {{\eta }_{\text{W}}} $ is the detection threshold, and the expression for $ {{\mathcal{P}}_{1}}\left( \beta {{\eta }_{\text{W}}} \right) $ is obtained from \eqref{eq82} as
\begin{equation}
	\label{eq103}
	{{\mathcal{P}}_{1}}\left( \beta {{\eta }_{\text{W}}} \right)=\sum\limits_{k=0}^{L-N-1}{\text{C}_{L-N+p-1}^{k+p}}\frac{{{\left( \beta {{\eta }_{\text{W}}} \right)}^{k+p}}\text{IG}\left( \frac{{{\rho }_{\theta }}\beta }{1+\beta {{\eta }_{\text{W}}}} \right)}{{{\left( 1+\beta {{\eta }_{\text{W}}} \right)}^{L-N+p-1}}}.
\end{equation}
Using a similar treatment to \eqref{eq102}, the PFA of the SAMF-NMC can be expressed as
\begin{equation}
	\label{eq104}
	\text{PF}{{\text{A}}_{\text{SAMF-NMC}}}=1-\int_{0}^{1}{{{\mathcal{P}}_{0}}\left( \beta {{\eta }_{\text{W}}} \right){{f}_{0}}\left( \beta  \right)\text{d}\beta }.
\end{equation}

Note that the expressions for the PDs of the three proposed detectors and the PFAs of the SRao-NMC and SAMF-NMC are analytical, while the expression for the PFA of the SGLRT-NMC is in closed-form. These analytical expressions can be efficiently computed using standard numerical integration techniques. 

The analytical results demonstrate that the performance of the proposed detectors is affected by factors such as SCR, signal mismatch $\cos^2\theta$, the signal subspace dimension $p$, and the number of training data $L$. In the following, the validity of the analytical expressions will be verified using both simulated and measured data, and we attempt to quantify how each factor influences the detection performance.

We would like to emphasize that the analytical PFA expressions of the proposed detectors play a crucial role in practical system implementation. They can be directly used to calculate the detection threshold accurately. This eliminates the need for time-consuming and computationally expensive Monte Carlo simulations, thereby improving the efficiency of system parameter design and optimization.

Meanwhile, the analytical PD expressions of the detectors provide significant practical guidance. They enable the precise prediction of detection performance under different SCRs. Furthermore, from these analytical expressions, the key physical quantities that affect the detection performance, e.g., signal subspace dimension, clutter covariance characteristics, and SCR, can be clearly identified. This insight not only helps in evaluating the detection capability of the proposed detectors under various practical conditions but also provides direct guidance for formulating effective measures to improve detection performance, and even offers valuable references for the overall radar system design process.

\section{Numerical Examples}
This section first presents a practical radar sensing application context to explicitly link the proposed detectors to real-world implementation, followed by performance evaluations using simulated data and measured data in sequence.
\subsection{Practical Radar Sensing Application Context}
The three proposed multichannel adaptive detectors are generalizable to spatial, temporal, and space-time multi-channel scenarios. To explicitly illustrate their practical implementation and address the gap between analytical work and real-world applications, we take X-band maritime surveillance pulse-Doppler radar as a concrete case study, focusing on temporal Doppler processing for target detection.

We consider a maritime surveillance X-band radar with system parameters to establish a direct mapping between theoretical variables and engineering practice.
The center frequency is 9.4 GHz, which leads to the
wavelength $\lambda = c/f_{c} \approx 0.032$ m. The bandwidth (BW) is
6 MHz, which achieves the range
resolution$\Delta R = \frac{c}{2  \text{BW}} = 25\ m$. The pulse
repetition frequency (PRF) is 1 kHz, and the coherent processing
interval (CPI) is 12 ms. The Doppler resolution, which is minimum
distinguishable Doppler frequency difference, can be calculated as
$\Delta f_{d} = 1/T_{\text{CPI}} = 1/(NT_{\text{r}}) = f_{\text{r}}/N \approx 83.33$
Hz.

The core mapping between theoretical parameters and the radar system is
clarified as follows:

1) Observation dimension $N$. This is the number of pulses in one CPI,
which is calculated as $N = \text{PRF} \times \text{CPI} = 12$. Hence,
a $12 \times 1$ test vector $\mathbf{x}$ is constructed from 12
consecutive radar pulses. This vector captures the temporal Doppler
characteristics of targets and clutter.

2) Signal subspace dimension $p$. This can arise from at least either of the following two scenarios.

${\scriptstyle\bullet}$  Scenario 1: Multiple targets in the same resolution cell. 3
ships within a 25 m range cell, each with distinct radial velocities (3
m/s, 5 m/s, 7 m/s), leading to $p = 3$. The signal matrix
$\mathbf{A}$ is constructed using steering vectors corresponding to
their respective Doppler frequencies ($f_{d} = 2v_{r}/\lambda$). Precisely, $\mathbf{A}$ has the form 
\begin{equation}
\label{}
\mathbf{A}=[\mathbf{a}(f_{d_1}),\mathbf{a}(f_{d_2}),\mathbf{a}(f_{d_3})],
\end{equation}
with
\begin{equation}
\label{}
\mathbf{a}(f_{d_1})=[1,\text{e}^{-j2\pi f_{d_i}T_r},\cdots,\text{e}^{-j2\pi(N-1)f_{d_i}T_r} ]^\text{T},
\end{equation}
 $f_{d_1}=187.5 \text{Hz}$, $f_{d_2}=312.5 \text{Hz}$ , and $f_{d_3}=437.5 \text{Hz}$.

${\scriptstyle\bullet}$ Scenario 2: Target smearing over a long CPI: A maneuvering
ship with average radial velocity $v_{r} = 5$ m/s and velocity
variation range $\Delta v_{r} = 4$ m/s (e.g., accelerating from 3 m/s
to 7 m/s) during the 12 ms CPI. This velocity spread induces a Doppler
frequency spread of
$\Delta f_{d}^{\text{sprd}} = 2\Delta v_{r}/\lambda \approx 250$ Hz,
which covers $\Delta f_{d}^{\text{sprd}}/\Delta f_{d} \approx 3$
Doppler resolution intervals. The target echo is thus smeared across 3
Doppler bins, forming a 3-dimensional signal subspace i.e., $p = 3$.
The signal matrix $\mathbf{A}$ includes steering vectors for the
Doppler frequencies corresponding to the minimum
($v_{r} - \Delta v_{r}/2 =$3m/s and $f_{d} \approx 187.5$ Hz),
average (5 m/s and $f_{d} \approx 312.5$ Hz), and maximum
($v_{r} + \Delta v_{r}/2 =$7 m/s and $f_{d} \approx 437.5$ Hz)
velocities.

3) The number of training data $L$. This is the number of range cells
in the vicinity of the test data ensure reliable estimation of
clutter mean and covariance. $L = 2N$ corresponds to using data within
a range of $L \cdot \Delta R = 24\times25 = 600m$ as training data.

4) Nonzero-mean clutter sources. In this X-band scenario, nonzero-mean
clutter can originate from discrete interferers (floating debris, birds,
or small boats outside the main beam), sidelobe clutter from coastal
terrain (cliffs, buildings), or illumination variations (uneven sea
surface reflectivity under sunlight).

%
%

In the following numerical examples, we will adopt similar parameter settings to evaluate the detection performance under these practical conditions.

In order to fully evaluate the detection performance, a comparison is made with the SGLRT\cite{RaghavanPulsone96}, SAMF\cite{LiuZhang12b}, and SRao\cite{LiuXie14b}, which do not take into account the nonzero-mean value of the clutter, and the detection statistics of the three detectors are as follows
\begin{equation}
	\label{eq105}
	t_{\text{SGLRT}} =
\frac{\tilde{\mathbf{x}}^{\text{H}}\mathbf{P}_{\tilde{\mathbf{A}}}\tilde{\mathbf{x}}} {1+\tilde{\mathbf{x}}^{\text{H}}\mathbf{P}_{\tilde{\mathbf{A}}}^\bot\tilde{\mathbf{x}}}
\end{equation}
\begin{equation}
	\label{eq106}
	t_{\text{SAMF}}=
{\tilde{\mathbf{x}}^{\text{H}}\mathbf{P}_{\tilde{\mathbf{A}}}\tilde{\mathbf{x}}} ,
\end{equation}
\begin{equation}
	\label{eq107}
	\begin{aligned}
	t_{\text{SRao}}=
\frac{\tilde{\mathbf{x}}^{\text{H}}\mathbf{P}_{\tilde{\mathbf{A}}}\tilde{\mathbf{x}}} {(1+\tilde{\mathbf{x}}^{\text{H}}\mathbf{P}_{\tilde{\mathbf{A}}}^\bot\tilde{\mathbf{x}})(1+\tilde{\mathbf{x}}^{\text{H}}\tilde{\mathbf{x}} )}
\end{aligned}
\end{equation}
where $\tilde{\mathbf{x}}=\mathbf{S}^{-\frac{1}{2}}\mathbf{x}$, $\tilde{\mathbf{A}}=\mathbf{S}^{-\frac{1}{2}}\mathbf{A}$,
$\mathbf{P}_{\tilde{\mathbf{A}}} =\tilde{\mathbf{A}}{(\tilde{\mathbf{A}}^{\text{H}}\tilde{\mathbf{A}} )}^{-1}\tilde{\mathbf{A}}^{\text{H}}$, $\mathbf{P}_{\tilde{\mathbf{A}}}^\bot=\mathbf{I}_N-\mathbf{P}_{\tilde{\mathbf{A}}} $, and
$ \mathbf{S}=\sum\limits_{{\ell}=1}^{L}{{{\mathbf{x}}_{\ell}}\mathbf{x}_{\ell}^{\text{H}}} $.

To better illustrate the distinctive features of the proposed detectors, the SGLRT-NMC is compared with two detectors. One is designed for rank-1 signals in nonzero-mean clutter, referred to as GLRT in nonzero-mean clutter (GLRT-NMC)\footnote{The real-valued version of the GLRT-NMC is given in \cite{OlivierFrancois21SP}.}, while the other is tailored for subspace signals in zero-mean clutter, i.e., the SGLRT in \eqref{eq105}. A detailed comparison is summarized in Table \ref{tab2}, which contrasts key characteristics including the clutter model, test statistic, core processing logic, statistical distribution under hypothesis $\text{H}_1$, and the output SCR, where $\dot{\mathbf{z}}=\mathbf{S}_2^{-1/2} \mathbf{z}$, $\dot{\mathbf{A}}=\mathbf{S}_2^{-1/2} \mathbf{A}$, $\dot{\mathbf{a}}=\mathbf{S}_2^{-1/2} \mathbf{a}$, $\mathbf{a}$ is the signal steering vector for rank-one signals, $\mathbf{p}_0$ denotes the actual signal steering vector.

Table \ref{tab2} clearly illustrates the core similarities and differences in the test statistics and characteristics between the proposed detectors and the two conventional detectors. While the test statistics of all three detectors share a similar structure, their fundamental distinction lies in the signal and clutter models they address. The GLRT-NMC is designed for rank-1 signals in nonzero-mean clutter, the SGLRT is proposed for subspace signals in zero-mean clutter, whereas our proposed SGLRT-NMC innovates by successfully integrating a subspace signal model with a nonzero-mean clutter environment. A distinction that is also directly reflected in their respective parameters of the complex F-distribution and complex Beta distribution.

\begin{table*}[htbp]
	\centering
	\caption{Comparison of detectors: Test statistic and key assumptions}
	\label{tab2}
	\setlength{\tabcolsep}{6pt}
	\renewcommand{\arraystretch}{1.2}
	\begin{tabularx}{\linewidth}{@{}>{\raggedright\arraybackslash}p{2.5cm}
			>{\centering\arraybackslash$}p{2cm}<{$}
			>{\centering\arraybackslash}X
			>{\centering\arraybackslash$}X<{$}
			>{\centering\arraybackslash$}X<{$}@{}}
		\toprule
		\multicolumn{1}{c}{\makecell{Detector}} &
		\multicolumn{1}{c}{\makecell{Test-Statistic}} &
		\multicolumn{1}{c}{\makecell{Key Assumptions}} &
		\multicolumn{1}{c}{\makecell{Statistical Distribution ($\text{H}_1$)}} &
		\multicolumn{1}{c}{\makecell{Output SCR}} \\
		\cmidrule(lr){1-5}

		\makecell{SGLRT-NMC} &
		\dfrac{\dot{\mathbf{z}}^{H}\mathbf{P}_{\mathbf{A}}\dot{\mathbf{z}}}{1+\dot{\mathbf{z}}^{H}\dot{\mathbf{z}}} &
		\makecell{Subspace signal, \\ nonzero-mean clutter} &
		\begin{aligned}
			&t_{\text{SGLRT-NMC}}[\beta; \text{H}_1] \sim\mathcal{CF}_{p,L-N}(\beta\rho_\theta) \\
			&\beta\mid \text{H}_1 \sim \mathcal{CB}_{L-N+p,N-p}(\delta^2)
		\end{aligned} &
		\dfrac{L}{L+1} \mathbf{p}_0^{H}\mathbf{R}^{-1}\mathbf{p}_0 \\
		
		\addlinespace[1.2em]
		
		\makecell{GLRT-NMC} &
		\dfrac{|\dot{\mathbf{a}}^{H}\dot{\mathbf{z}}|^{2}}{\dot{\mathbf{a}}^{H}\dot{\mathbf{a}}(1+\dot{\mathbf{z}}^{H}\dot{\mathbf{z}})} &  
		\makecell{Rank-1 signal, \\ nonzero-mean clutter} &
		\begin{aligned}
			&t_{\text{GLRT-NMC}}[\beta; \text{H}_1] \sim \mathcal{CF}_{1,L-N}(\beta\rho_\theta) \\
			&\beta\mid \text{H}_1 \sim \mathcal{CB}_{L-N+1,N-1}(\delta^2)
		\end{aligned} &
		\dfrac{L}{L+1} \mathbf{p}_0^{H}\mathbf{R}^{-1}\mathbf{p}_0 \\
		
		\addlinespace[1.2em]

		\makecell{SGLRT} &
		\dfrac{\tilde{\mathbf{x}}^{H}\mathbf{P}_{\mathbf{A}}\tilde{\mathbf{x}}}{1+\tilde{\mathbf{x}}^{H}\tilde{\mathbf{x}}} &
		\makecell{Subspace signal, \\ zero-mean clutter} &
		\begin{aligned}
			&t_{\text{SGLRT}}[\beta; \text{H}_1] \sim\mathcal{CF}_{p,L-N+1}(\beta\rho_\theta) \\
			&\beta\mid \text{H}_1 \sim\mathcal{CB}_{L-N+p+1,N-p+1}(\delta^2)
		\end{aligned} &
		\mathbf{p}_0^{H}\mathbf{R}^{-1}\mathbf{p}_0 \\
		
		\bottomrule
	\end{tabularx}
\end{table*}

\subsection{Performance Evaluations Using Simulated Data}
Although the nonzero mean of clutter has no impact on the detection performance of the proposed detectors, it does influence the performance of the SGLRT, SAMF, and SRao\footnote{The statistical characterization of the  SGLRT, SAMF, and SRao in a nonzero-mean clutter has so far been an unresolved problem. We find that the metrics in \eqref{eq108} and \eqref{eq109} affect the performance of the three detectors.}, for which two metrics are introduced. One is the amplitude power of the nonzero-mean of the clutter
\begin{equation}
	\label{eq108}
	\xi ={{\bm{\mu} }^{\text{H}}}{{\mathbf{R}}^{-1}}\bm{\mu} .
\end{equation}
The other is the squared cosine of the angle of the signal $ {{\mathbf{p}}}_{0} $ and the nonzero mean of the clutter $ \bm{\mu} $ in whitened space, i.e.,
\begin{equation}
	\label{eq109}
	{{\cos }^{2}}\phi =\frac{|\mathbf{p}_{0}^{\text{H}}{{\mathbf{R}}^{-1}}\bm{\mu} {{|}^{2}}}{\mathbf{p}_{0}^{\text{H}}{{\mathbf{R}}^{-1}}{{\mathbf{p}}_{0}}\cdot {{\bm{\mu} }^{\text{H}}}{{\mathbf{R}}^{-1}}\bm{\mu} }.
\end{equation}

The $({{n}_{1}},{{n}_{2}})$th element of the covariance matrix is set to $\mathbf{R}({{n}_{1}},{{n}_{2}})={{\varepsilon }^{|{{n}_{1}}-{{n}_{2}}|}}$, ${{n}_{1}},{{n}_{2}}=1,2,\cdots ,N$.
The relevant parameters are set as follows: $N=12$, $\varepsilon=0.95$, and $\text{PFA}={{10}^{-3}}$.
The acquisition of a PD value is conducted through $10^4$ Monte Carlo simulations, while determining a detection threshold requires $10^5$ simulations.

In Appendix \ref{app:ParameterGeneration}, we detail the procedure for generating the signal matrix $\mathbf{A}$, actual signal steering vector $\mathbf{p}_0$, clutter's nonzero mean $\bm{\mu}$, ensuring they satisfy specific preset values of $\cos^2{\theta}$ defined in \eqref{eq74}, SCR defined in \eqref{eq75}, $\xi$ defined in \eqref{eq108}, $\cos^2{\phi}$ defined in \eqref{eq109}.

\begin{figure}
	\centerline{\includegraphics[width=0.37\textwidth]{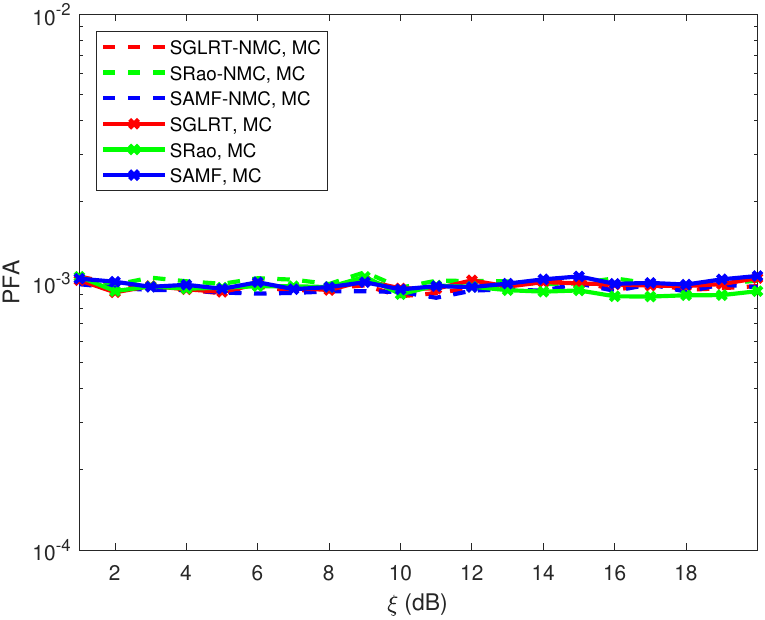}}
	\caption{PFA versus the energy of $ \xi $ with simulated data. ($N{=12}$, $p=3$, $L=2N$, $ {{\cos }^{2}}\phi =0.3 $)}
	\label{1_New}
\end{figure}

\begin{figure}
	\centerline{\includegraphics[width=0.37\textwidth]{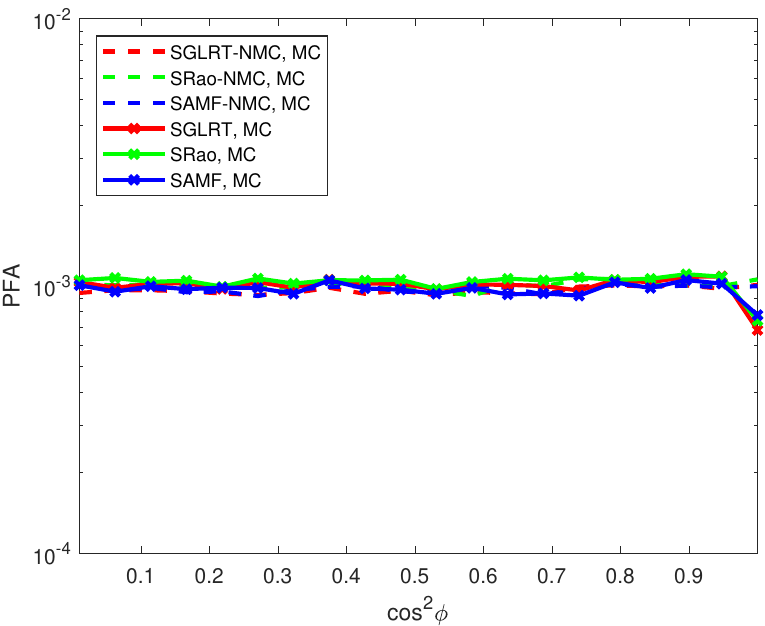}}
	\caption{PFA versus $\text{cos}^2\phi$ with simulated data. ($N{=12}$, $p=3$, $L=2N$, $\xi =35~\text{dB}$)}
	\label{2_New}
\end{figure}

Figs. \ref{1_New} and \ref{2_New} illustrate the variation of PFA with respect to $\xi$ and $\cos^2\phi$, respectively. It can be seen that the detectors proposed in this paper maintain CFAR properties against both $\xi$ and $\cos^2\phi$. In contrast, conventional detectors exhibit CFAR behavior with regard to $\xi$, yet their PFA varies as $\cos^2\phi$ approaches 1, indicating a lack of strict CFAR performance in terms of $\cos^2\phi$. 

\begin{figure}
	\centerline{\includegraphics[width=0.37\textwidth]{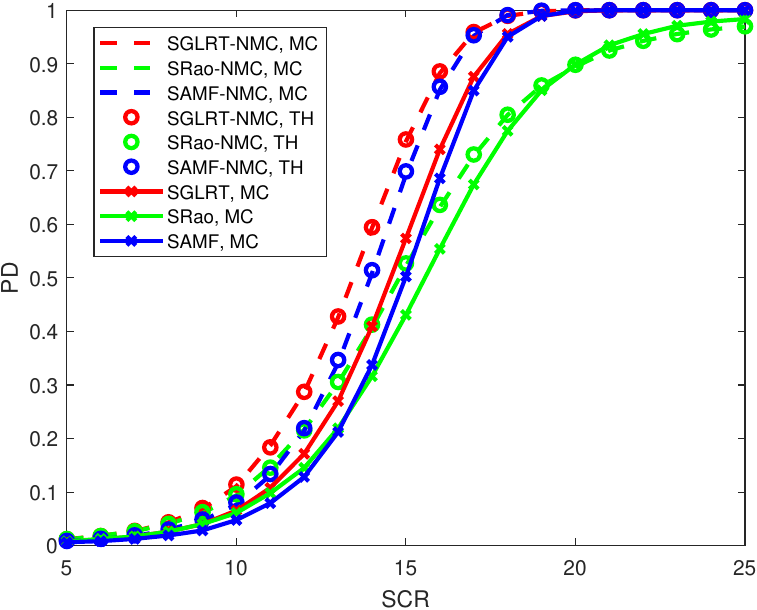}}
	\caption{PD versus SCR with simulated data. ($N{=12}$, $p=3$, $L=2N$, $ {{\cos }^{2}}\phi =0.3 $, $ \xi =35~\text{dB} $, $ {{\cos }^{2}}\theta =1 $)}
	\label{1}
\end{figure}

Fig. \ref{1} presents the theoretical values (``TH'') and Monte Carlo simulation results (``MC'') of the PDs versus SCR for the proposed SGLRT-NMC, SRao-NMC, and SAMF-NMC, with comparative results from conventional SGLRT, SAMF, and SRao. As evidenced by the result, the theoretical detection performance and Monte Carlo simulation results of the three proposed detectors exhibit highly consistent agreement, thereby validating the statistical correctness of derived statistical performance.
The results in Fig. \ref{1} also illustrate the improvement of detection performance of the proposed detectors, compared to the traditional ones. Traditional detectors SGLRT, SAMF, and SRao assume zero-mean clutter and use the standard SCM ${{\mathbf{S}}_{*}}=\sum\nolimits_{{\ell}=1}^{L}{ {\mathbf{x}}_{\ell}{\mathbf{x}}_{\ell}^{\text{H}}}$ for whitening. In nonzero-mean clutter, ${{\mathbf{S}}_{*}}$ becomes a biased estimate of the true covariance ${{\mathbf{R}}}$, as it incorporates the clutter mean power. This bias distorts the whitening process, leading to mismatched filtering and subsequent performance degradation.

\begin{figure}
	\centerline{\includegraphics[width=0.37\textwidth]{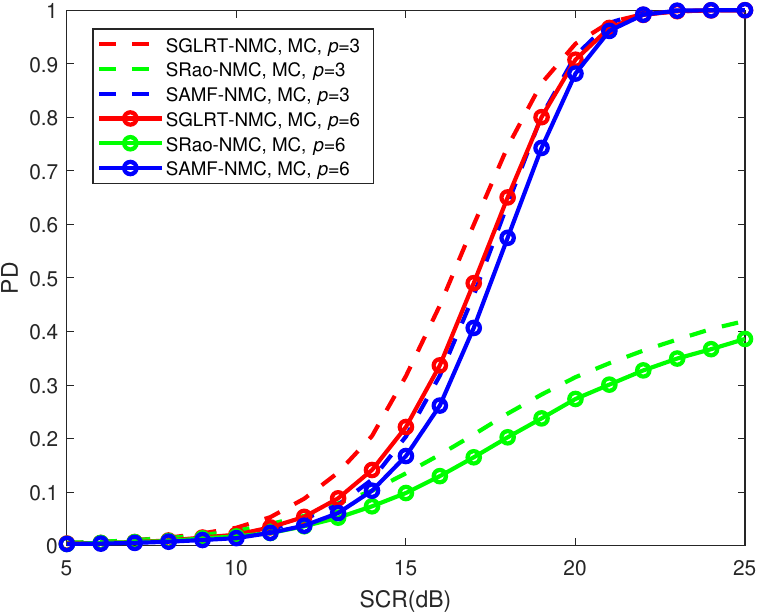}}
	\caption{PD versus SCR for different $p$ with simulated data.
		($N{=12}$, $L=1.5N$, $ {{\cos }^{2}}\phi =0.3$, $ \xi =35~\text{dB} $, $p=3$ and $p=6$, $ {{\cos }^{2}}\theta =1 $)}
	\label{2}
\end{figure}

Fig. \ref{2} illustrates the PD versus SCR of the detectors with reduced training data size.
A comparison of the results in Figs. \ref{1} and \ref{2} reveals that as the number of training data decreases, the PD of each detector decreases. This is because the estimation accuracy of unknown parameters such as the mean and covariance matrix of the clutter is affected by the count of the training data. The fewer the training samples, the lower the estimation accuracy. Hence, the corresponding PD is also lower.
Fig. \ref{2} also indicates that the detection performance of the detectors deteriorates with increasing subspace dimension $p$. This degradation occurs because larger values of $p$ introduce greater uncertainty in target orientation, thereby reducing detection capability.

\begin{figure}
	\centerline{\includegraphics[width=0.37\textwidth]{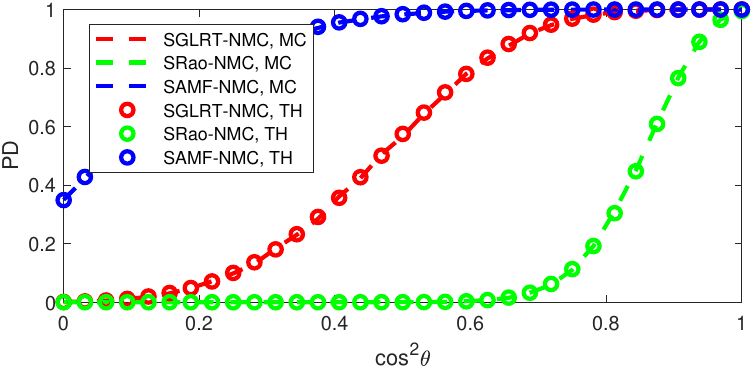}}
	\caption{PD versus $\text{cos}^2\theta$ with simulated data. ($N{=12}$, $p=3$, $L=2.5N$, $ {{\cos }^{2}}\phi =0.3 $, $ \text{SCR~=~20~dB} $, $ \xi =35~\text{dB} $)}
	\label{3}
\end{figure}

Fig. \ref{3} presents both theoretical and simulation results of the PD versus mismatch level for the proposed detectors. The results demonstrate consistent agreement between theoretical predictions (``TH'') and simulated outcomes (``MC'') across all three detectors. 
In addition, the PD of the SAMF-NMC decreases more slowly as $ {{\cos }^{2}}\theta $ decreases, while the PD of the SRao-NMC decreases sharply as $ {{\cos }^{2}}\theta $ decreases.Thus the SAMF-NMC has good robustness to mismatch signals, the SRao-NMC has strong mismatch selectivity, and the SGLRT-NMC is in between the above two detectors.

\begin{figure}
	\centerline{\includegraphics[width=0.37\textwidth]{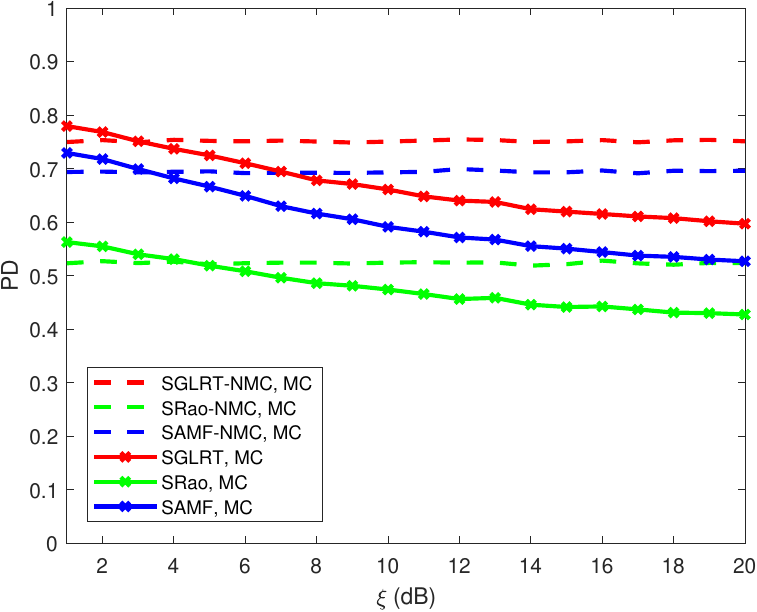}}
	\caption{PD versus the energy of $ \xi $ with simulated data. ($N{=12}$, $p=3$, $L=2N$, $ {{\cos }^{2}}\phi =0.3 $, $ \text{SCR~=~15~dB} $, $ {{\cos }^{2}}\theta =1 $)}
	\label{4}
\end{figure}

Fig. \ref{4} demonstrates the variation in detection performance across different clutter power levels for each detector. As shown, the PDs of the proposed detectors (SGLRT-NMC, SRao-NMC, and SAMF-NMC) remain unaffected by variations in nonzero-mean clutter power. In contrast, the PDs of conventional detectors (SGLRT, SAMF, and SRao) decrease significantly as the nonzero mean clutter power increases. In other words, the proposed detectors effectively suppress nonzero-mean clutter, while the conventional detectors fail.

\begin{figure}
	\centerline{\includegraphics[width=0.37\textwidth]{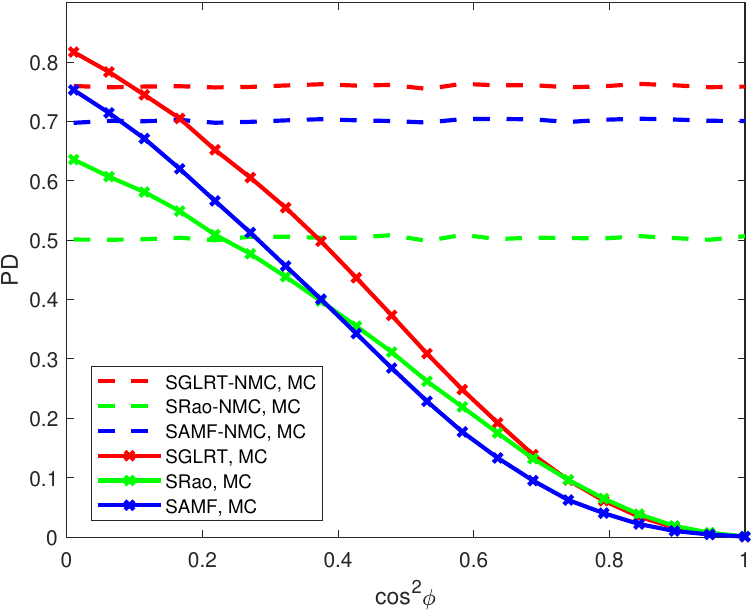}}
	\caption{PD versus $\text{cos}^2\phi$ with simulated data. ($N{=12}$, $p=3$, $L=2N$, $\xi =35~\text{dB}$, $\text{SCR~=~15~dB}$, ${{\cos }^{2}}\theta =1$)}
	\label{5}
\end{figure}

Fig. \ref{5} presents the PD versus $\cos^2 \phi$ between the clutter's nonzero mean $\bm{\mu}$ and the signal component ${{\mathbf{p}}_{0}}$ in the whitened space for all detectors. As evidenced by the figure, the proposed detectors maintain stable PDs regardless of the variation of $\cos^2 \phi$, demonstrating effective suppression capability against nonzero-mean clutter. In contrast, the PDs of conventional detectors degrade significantly as $\cos^2 \phi$ decreases, revealing their susceptibility to nonzero-mean clutter interference.

\subsection{Performance Evaluations Using Measured Data}
The experimental validation employed three measured datasets: 1) the ``\seqsplit{20221112175025\_stare\_VV}" dataset from the sea-detecting radar data-sharing program (SDRDSP) published by Naval Aviation University \cite{GuanLiu23}, featuring maritime surveillance X-band clutter with non-stationary characteristics;
2) the ``\seqsplit{19980205\_184733\_ANTSTEP}" dataset from the ice multiparameter imaging X-band (IPIX) radar \cite{TangWang20SPL}, containing heterogeneous ice/sea clutter; and
3) the ``\seqsplit{UHF\_08\_SS2\_HH\_BW2.5M\_PRF1000}" dataset released by the 22nd institute of China electronics technology group corporation (CETC), comprising UHF-band clutter with high-resolution spatial variations.
The \seqsplit{20221112175025\_stare\_VV} data is the sea clutter data, fitted well by a compound-Gaussian distribution, collected by the VV polarization radar with 2000 Hz pulse repetition frequency (PRF) at sea state level 5. The data contains three pulse emission modes, of which amplitude\_complex\_T1 is the sea clutter data collected in mode 1, containing 950 range cells, each containing 67584 pulse data. 
The dataset \seqsplit{19980205\_184733\_ANTSTEP} was collected by the VV polarization radar with 1000 Hz PRF and 5.5MHz bandwidth (BW) in 1998 on the shore of Lake Ontario in Grimsby, Canada, and the collected data contain 28 range cells, each containing 60000 pulses, and also follow compound-Gaussian distributions \cite{Gini2000PerformanceAO}.
The \seqsplit{UHF\_08\_SS2\_HH\_BW2.5M\_PRF1000} dataset is the HH-polarized sea clutter data with 1000 Hz PRF and 2.5MHz BW under secondary sea state, which contains 56 range cells, each containing 59860 pulses data, and obeys Gaussian distribution.
For the sake of brevity in writing, the ``\seqsplit{20221112175025\_stare\_VV}'' is abbreviated as the ``\seqsplit{SDRDSP\_VV\_X\_CG}'', the ``\seqsplit{19980205\_184733\_ANTSTEP}'' is abbreviated as the ``\seqsplit{IPIX\_VV\_X\_CG}", and the ``\seqsplit{UHF\_08\_SS2\_HH\_BW2.5M\_PRF1000}" is abbreviated as the ``\seqsplit{CETC\_HH\_UHF\_G}''. Some available key parameters for the measured datasets are summarized in Table \ref{tab3}.

\begin{table*}[htbp]
	\centering
	\caption{Key information of measured datasets}
	\begin{tabular}{@{}cccccc@{}}
		\toprule
		Short ID & Polarization Mode & Frequency Band & BW & PRF & Original Clutter Type\\
		\midrule
		SDRDSP\_VV\_X\_CG & VV & X & —— & 2 KHz & Compound-Gaussian \\
		IPIX\_VV\_X\_CG  & VV & X & 5.5 MHz & 1 KHz & Compound-Gaussian \\
		CETC\_HH\_UHF\_G  & HH & UHF & 2.5 MHz & 1 KHz & Gaussian \\
		\bottomrule
	\end{tabular}
	\label{tab3}
\end{table*}

The scenarios addressed in this paper are Gaussian distributed clutter, so the SDRDSP dataset and the IPIX dataset need to be Gaussianized before carrying out experiments with real data. Otherwise, the detectors would suffer from performance degradation, as the CFAR property would be lost. In addition, the Gaussianized SDRDSP data and IPIX data, as well as the CETC data, can be also standardized in order to ensure that all data can be analyzed and compared under a uniform metric. Standardization refers to the transformation of data so that it conforms to a standard complex normal distribution with a standard deviation of 1.
All three measured datasets contain clutter-only data. This allows for the controlled injection of synthetic targets with known parameters. 
For each of the three measured data described above, the range cells in the $14-L/2$ to $14+L/2$ range and the pulse data in the 10,000-th to 50,000-th range are selected for subsequent preprocessing and detection performance analysis, where the target is assumed to be located in the 14th range cell. Meanwhile, due to the limited data available, the PFA is set to $\text{PFA}={{10}^{-2}}$ in the real data validation experiment. The intercepted clutter data is expressed as
\begin{equation}
	\label{eq111}
	\mathbf{C}=\left[ {{\mathbf{c}}_{14-L/2}},{{\mathbf{c}}_{15-L/2}},\cdots ,{{\mathbf{c}}_{14+L/2}} \right],
\end{equation}
where ${{\mathbf{c}}_{i}} = [{{c}_{i,m}}, {{c}_{i,m+1}}, \cdots, {{c}_{i,m+{{L}_{c}}-1}}]^{\text{T}}$ represents the measured data vector from the $i$-th range cell, with ${{c}_{i,j}}$ denoting the measured data corresponding to the $j$-th pulse in the $i$-th range cell. Here, $m=10000$ indicates the starting pulse index, and ${{L}_{c}}=40001$ specifies the number of pulses extracted from each individual range cell.

Compound-Gaussian clutter can be modeled as
\begin{equation}
	\label{eq112}
	c_{i,j} = \sqrt{\tau_{i,j}} \eta_{i,j}, \quad
	\begin{aligned}
		&i = 14-L/2,15-L/2,\dots,14+L/2, \\
		&j = m,m+1,\dots,m+L_c-1,
	\end{aligned}
\end{equation}
where ${{\tau }_{i,j}}$ and ${{\eta }_{i,j}}$ correspond to the texture component and speckle component, respectively, with ${{\eta }_{i,j}}$ following a complex Gaussian distribution. The key to Gaussianization lies in estimating ${{\tau }_{i,j}}$ to eliminate the influence of the texture component on the data distribution. An effective estimator for ${{\tau }_{i,j}}$ is \cite{TangWang20SPL}
\begin{equation}
	\label{eq113}
	{{\hat{\tau }}_{i,j}}={{\left[ \sum\limits_{k=-K/2}^{K/2}{\frac{\left| {{c}_{i,j+k}} \right|}{ K+1  }\;} \right]}^{2}},
\end{equation}
where $K+1$ denotes the window size for estimating $ {{\tau }_{i,j}} $. The Gaussianized data can be represented as
\begin{equation}
	\label{eq114}
	\overline{c}_{i,j}= {c_{i,j}} \big/{\sqrt{\hat{\tau}_{i,j}}}.
\end{equation}
In this paper, $K=32$ is chosen for the Gaussianization of the measured data.

The data \seqsplit{IPIX\_VV\_X\_CG} and \seqsplit{IPIX\_VV\_X\_CG} are normalized together with the data \seqsplit{CETC\_HH\_UHF\_G}. The normalization of the data after Gaussianization of the first two data can be expressed as
\begin{equation}
	\label{eq115}
	{{\overset{\scriptscriptstyle\frown}{c}}_{i,j}}= {\sqrt{{\pi }/{2}\;}{{L}_{c}}{{{\bar{c}}}_{i,j}}} \Bigg/~ {\sum\limits_{j=m}^{m+{{L}_{c}}-1}{\left| {{{\bar{c}}}_{i,j}} \right|}}.
\end{equation}
The normalization of data \seqsplit{CETC\_HH\_UHF\_G} can be rewritten as
\begin{equation}
	\label{eq116}
	{{\overset{\scriptscriptstyle\frown}{c}}_{i,j}}= {\sqrt{{\pi }/{2}\;}{{L}_{c}}{{c}_{i,j}}} \bigg/ ~{\sum\limits_{j=m}^{m+{{L}_{c}}-1}{\left| {{c}_{i,j}} \right|}}.
\end{equation}

The standardized data can be approximated by a standard complex normal distribution, and the magnitude obeys a Rayleigh distribution with scale parameter $\sigma =1$, represented in PDF as
\begin{equation}
	\label{eq117}
	f\left( \left| {{{\overset{\scriptscriptstyle\frown}{c}}}_{i,j}} \right| \right)=\frac{{\left| {{{\overset{\scriptscriptstyle\frown}{c}}}_{i,j}} \right|}}{{{\sigma }^{2}}}\text{e}^ { -\frac{{{\left| {{{\overset{\scriptscriptstyle\frown}{c}}}_{i,j}} \right|}^{2}}}{2{{\sigma }^{2}}} }
={\left| {{{\overset{\scriptscriptstyle\frown}{c}}}_{i,j}} \right|}\text{e}^{-\frac{{{\left| {{{\overset{\scriptscriptstyle\frown}{c}}}_{i,j}} \right|}^{2}}}{2} },
\end{equation}
where $\sigma $ denotes the scale parameter of the Rayleigh distribution and also the standard deviation of its corresponding Gaussian distribution.

\begin{figure}[htbp]
	\centerline{\includegraphics[width=0.37\textwidth]{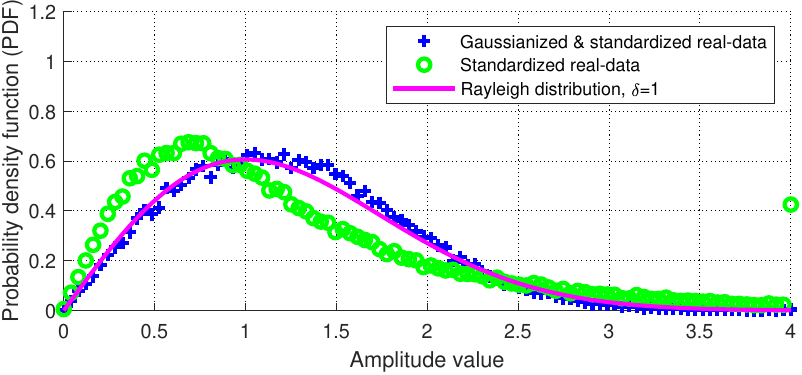}}
	\caption{Results of Rayleigh distribution fitting for the 14th range cell data in SDRDSP dataset 20221112175025\_stare\_VV after Gaussianization and Standardization processing.}
	\label{6}
\end{figure}

\begin{figure}[htbp]
	\centerline{\includegraphics[width=0.37\textwidth]{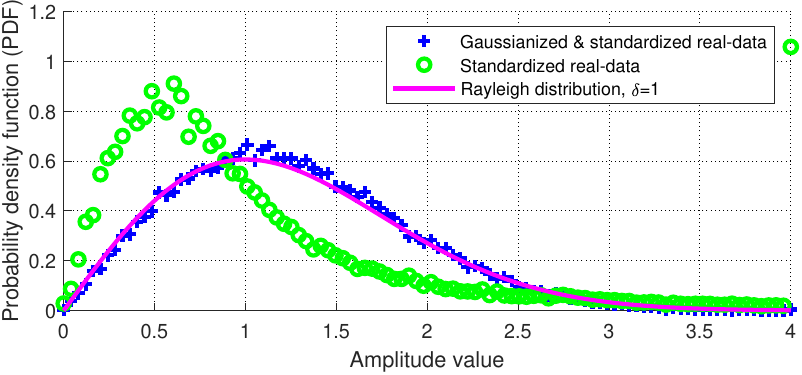}}
	\caption{Results of Rayleigh distribution fitting for the 14th range cell data in IPIX dataset IPIX\_VV\_X\_CG after Gaussianization and Standardization processing.}
	\label{7}
\end{figure}

\begin{figure}[htbp]
	\centerline{\includegraphics[width=0.37\textwidth]{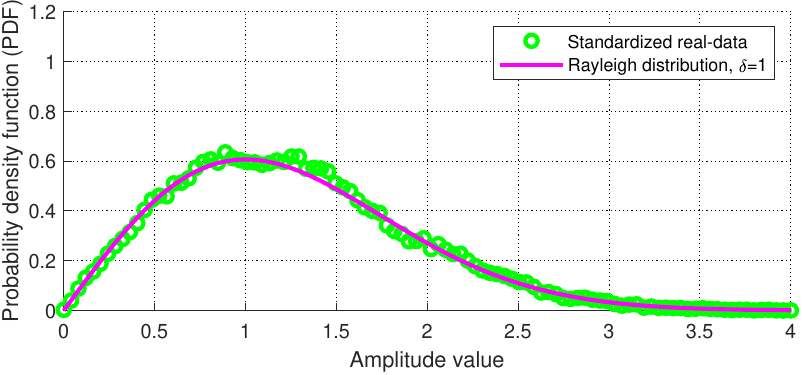}}
	\caption{Results of Rayleigh distribution fitting for the 14th range cell data in CETC dataset CETC\_HH\_UHF\_G after Gaussianization and Standardization processing.}
	\label{8}
\end{figure}

Figs. \ref{6} and \ref{7} respectively demonstrate the Rayleigh distribution fitting results for both Gaussian-normalized and merely standardized processing of the \seqsplit{IPIX\_VV\_X\_CG} and \seqsplit{IPIX\_VV\_X\_CG} data, while Fig. \ref{8} presents the fitting results exclusively for the standardized \seqsplit{CETC\_HH\_UHF\_G} data. The results demonstrate that for both the \seqsplit{IPIX\_VV\_X\_CG} data from the SDRDSP dataset and the \seqsplit{IPIX\_VV\_X\_CG} data from the IPIX radar, the amplitude modulus distributions deviate significantly from the Rayleigh distribution when only standardized processing is applied. However, after undergoing Gaussian normalization followed by standardization, these datasets show significantly improved conformity with the Rayleigh distribution. In contrast, the \seqsplit{CETC\_HH\_UHF\_G} data from the CETC achieves excellent Rayleigh distribution fitting with standardization processing alone.

\begin{figure}
	\centerline{\includegraphics[width=0.37\textwidth]{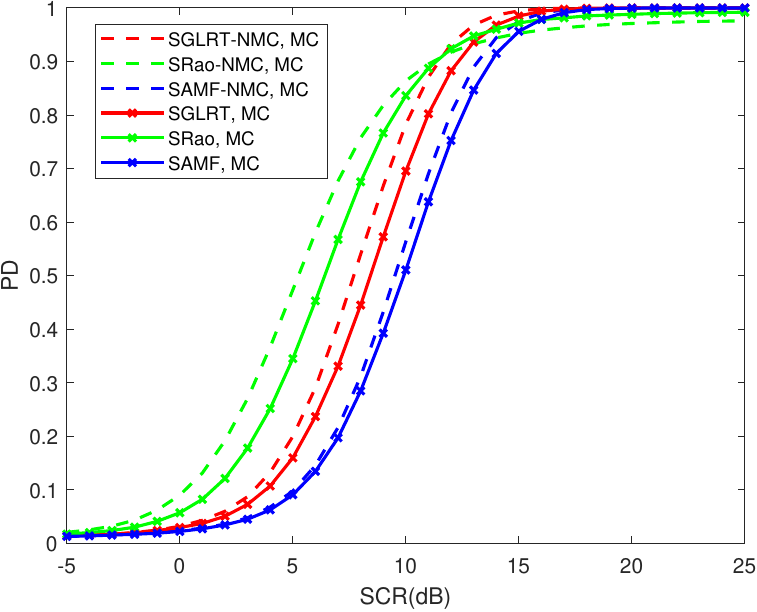}}
	\caption{PD versus SCR based on data 20221112175025\_stare\_VV.
		($N{=12}$, $p=3$, $L=2N$, $ {{\cos }^{2}}\theta =1 $)}
	\label{9}
\end{figure}

To further validate the non-zero mean property of the clutter, a targeted significance test is performed on its mean using the complex multivariate Hotelling’s T² test. The corresponding mathematical details are omitted for space limitation, and the interested reader is referred to \cite{Muirhead05} or other statistical texts.
Table \ref{tab1} presents the results of the Hotelling's T² test for the three datasets, with a significance level of $a$ = 0.001. The test result for SDRDSP fails to reject the null hypothesis that ``the population mean of the clutter is zero,'' whereas the results for both IPIX and CETC lead to the rejection of this null hypothesis. Therefore, the clutter mean in the IPIX and CETC datasets is statistically significantly nonzero\footnote{Despite the Hotelling's T² test failing to reject a zero clutter mean for the SDRDSP dataset, our detectors maintain superior performance as shown in Fig. \ref{9}. A possible explanation is that the estimated clutter mean is non-zero but statistically insignificant. Furthermore, the angle between the clutter mean vector and the target steering vector is crucial. Even a small mean can degrade conventional detectors if this angle is small, explaining the observed performance advantage.}.

\begin{table}[htbp]
	\centering
	\caption{Hotelling's T$^2$ test results on measured data}
	\begin{tabular}{@{}lcccccc@{}}
		\toprule
		Dataset & $L_r$ & $N$ & F-statistic & $a$ & $\text{F}$ critical value & Result \\
		\midrule
		SDRDSP & 3333 & 12 & 1.36 & 0.001 & 2.1371 & $\bm{\mu}=\bm{0}$ \\
		IPIX   & 3333 & 12 & 5945.88 & 0.001 & 2.1371 & $\bm{\mu}\neq\bm{0}$ \\
		CETC   & 3333 & 12 & 17.84 & 0.001 & 2.1371 & $\bm{\mu}\neq\bm{0}$ \\
		\bottomrule
	\end{tabular}
	\label{tab1}
\end{table}

The agreement with the Rayleigh distribution in Figs. \ref{6}-\ref{8} validates our preprocessing chain, confirming that the subsequent detection experiments are conducted on data that conforms to the assumed Gaussian model.

Since the covariance matrix of the real data is unknown and the statistical performance of the detectors depends on the true covariance matrix, in order to define the SCR based on the real data, the sample SCM composed of all Gaussianized/normalized measured data is used as an approximation of the true value, denoted as
\begin{equation}
	\label{eq118}
	\mathbf{\overset{\scriptscriptstyle\frown}{R}}=\frac{1}{(L+1){{L}_{r}}}\sum\limits_{i=14-L/2} ^{14+L/2}{\sum\limits_{s=1}^{{{L}_{r}}}{{{{\mathbf{\overset{\scriptscriptstyle\frown}{c}}}}_{i,s}} \mathbf{\overset{\scriptscriptstyle\frown}{c}}_{i,s}^{\text{H}}}},
\end{equation}
where ${{L}_{r}}=\left\lfloor {{{L}{s}}}/{N}\right\rfloor $ represents the number of constructed clutter vectors per range cell, with the notation $\left\lfloor \cdot \right\rfloor $ denoting the floor (downward rounding) operation, $ {{\mathbf{\overset{\scriptscriptstyle\frown}{c}}}_{i,s}}={{[{{\overset{\scriptscriptstyle\frown}{c}}_{i,m+N(s-1)}},{{\overset{\scriptscriptstyle\frown}{c}}_{i,m+N(s-1)+1}},\cdots ,{{\overset{\scriptscriptstyle\frown}{c}}_{i,m+Ns-1}}]}^{\text{T}}} $ represents the constructed clutter vector. Therefore, the SCR based on measured data is expressed as
\begin{equation}
	\label{eq119}
	\text{SCR}=
\frac{L}{L+1} \mathbf{p}_{0}^{\text{H}}{{\mathbf{\overset{\scriptscriptstyle\frown}{R}}}^{-1}} {{\mathbf{p}}_{0}}.
\end{equation}

\begin{figure}
	\centerline{\includegraphics[width=0.37\textwidth]{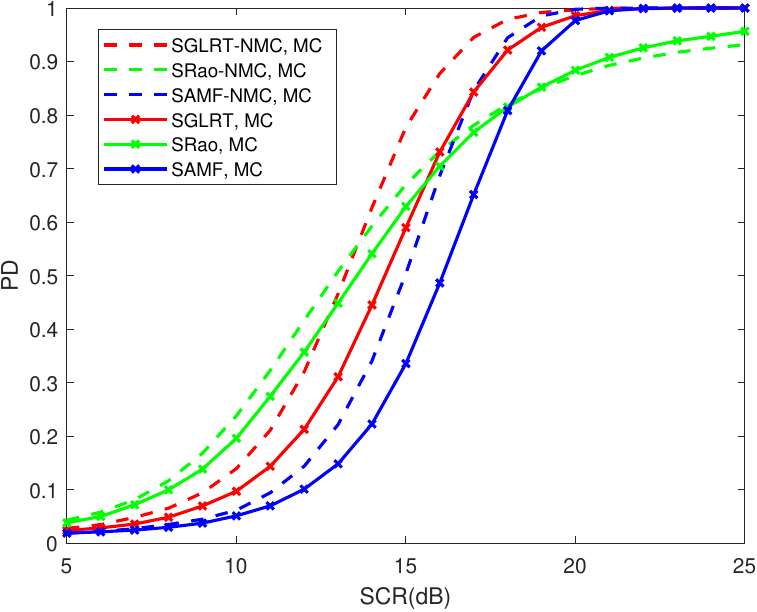}}
	\caption{PD versus SCR based on data IPIX\_VV\_X\_CG
		($N{=12}$, $p=3$, $L=2N$,  $ {{\cos }^{2}}\theta =1 $)}
	\label{10}
\end{figure}

\begin{figure}
	\centerline{\includegraphics[width=0.37\textwidth]{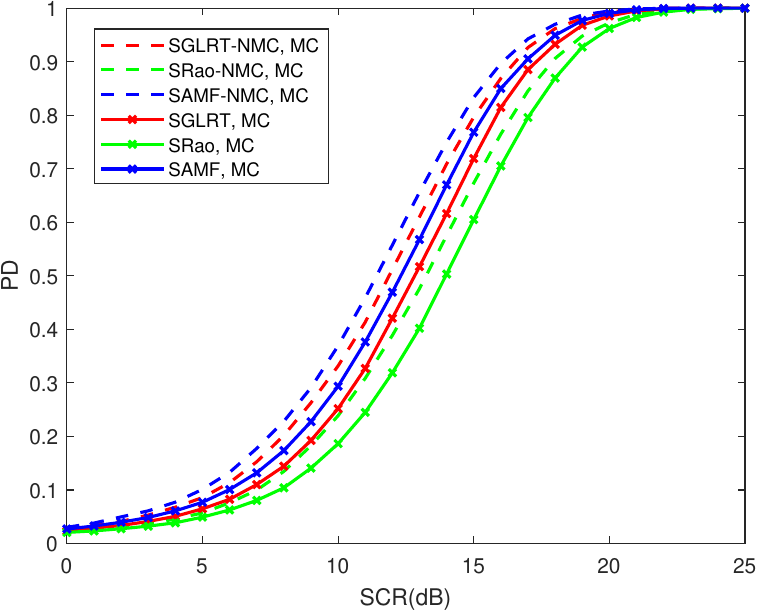}}
	\caption{PD versus SCR based on data CETC\_HH\_UHF\_G
		($N{=12}$, $p=3$, $L=2N$,  $ {{\cos }^{2}}\theta =1 $)}
	\label{11}
\end{figure}
\begin{figure}[htbp]
	\centering
	\begin{subfigure}[b]{0.37\textwidth}
		\includegraphics[width=\textwidth]{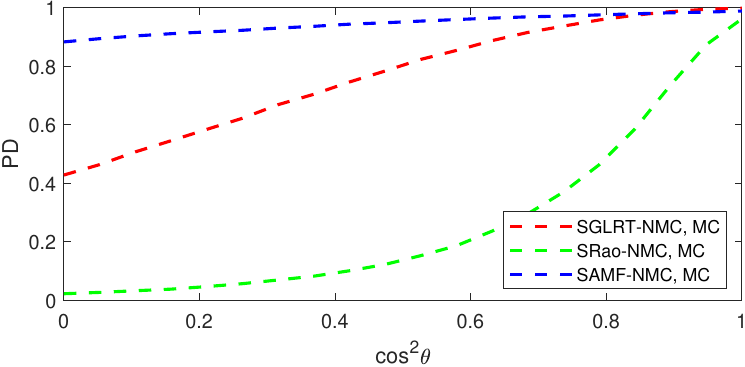}
		\caption{ 20221112175025\_stare\_VV, $N{=12}$, $p=3$, $L=2N$, $ \text{SCR=15dB} $}
		\label{fig12a}
	\end{subfigure}
	
	\vfill 
	
	\begin{subfigure}[b]{0.37\textwidth}
		\includegraphics[width=\textwidth]{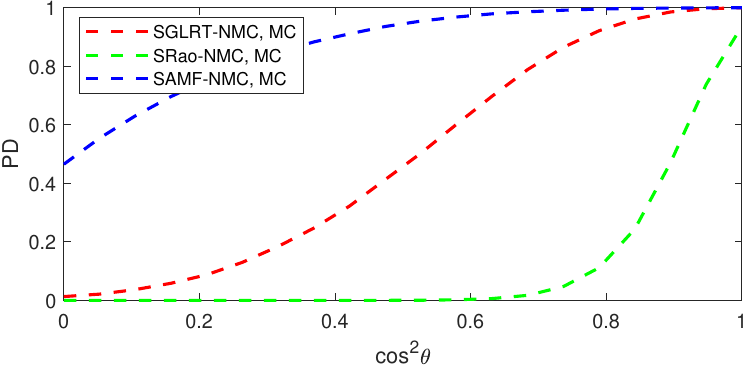}
		\caption{IPIX\_VV\_X\_CG, $N{=12}$, $p=3$, $L=2N$, $ \text{SCR=23dB} $}
		\label{fig12b}
	\end{subfigure}
	
	\vfill 
	
	\begin{subfigure}[b]{0.37\textwidth}
		\includegraphics[width=\textwidth]{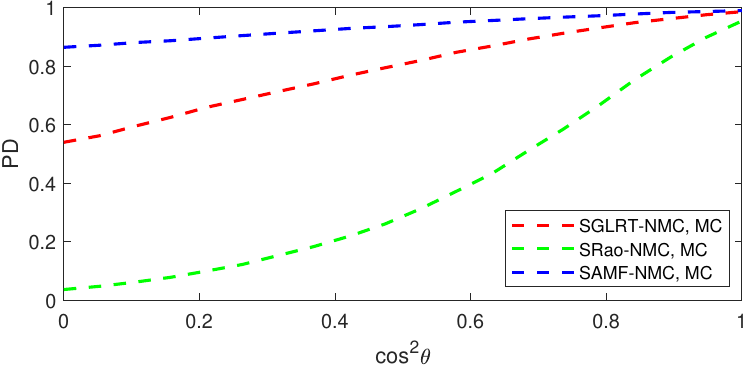}
		\caption{CETC\_HH\_UHF\_G, $N{=12}$, $p=2$, $L=2N$,  $ \text{SCR~=~20~dB} $}
		\label{fig12c}
	\end{subfigure}
	
	\caption{PD versus $\text{cos}^2\theta$ based on different datasets.}
	\label{12}
\end{figure}
Fig. \ref{9} presents the PD versus SCR for all detectors based on the \seqsplit{IPIX\_VV\_X\_CG} dataset from the SDRDSP measured dataset. It can be seen that the SGLRT-NMC always outperforms the SGLRT and SAMF; the SRao-NMC has the best detection performance for $\text{SCR}\leq 10\; \text{dB}$
; and the PD of the SAMF-NMC is slightly lower than that of the SGLRT, but higher than that of the SAMF.

Fig. \ref{10} presents the PD versus SCR for all detectors based on the \seqsplit{IPIX\_VV\_X\_CG} dataset from the IPIX radar measured data. It can be seen that the SGLRT-NMC consistently outperforms the SGLRT and SAMF; the SRao-NMC outperforms the SRao for $\text{SCR}\leq 17\; \text{dB}$
; and the SAMF-NMC has higher PD than the SAMF, has higher PD than the SGLRT for  $16~ \text{dB}\le \text{SCR}\leq 20\; \text{dB}$.

Fig. \ref{11} displays the PD versus SCR for all detectors based on the \seqsplit{CETC\_HH\_UHF\_G} dataset from the 22nd institute of CETC. It can be seen that the SAMF-NMC has the best detection performance, followed by the SGLRT-NMC; whereas the detection performance of the SGLRT and SAMF is in between that of the SGLRT-NMC and SRao-NMC; and the SRao has the worst detection performance.

Fig. \ref{12} presents the detection performance versus mismatch level for the detectors based on the above three datasets. Similar to the simulation data shown in Fig. \ref{3},
It is shown that the SAMF-NMC exhibits strong robustness, the SRao-NMC exhibits high selectivity, and the SGLRT-NMC is in between. The measured data results in Fig. \ref{12} corroborate the simulation-based findings in Fig. \ref{3}, confirming that the robustness/selectivity characteristics of the proposed detectors hold true in complex real-world environments.

In practice, the SGLRT-NMC is recommend in the absence of signal mismatch, whereas the SAMF-NMC and SRao-NMC are preferred for robust and selective detection, respectively, under mismatched conditions.

\section{Conclusion}

In this paper, we have addressed the problem of adaptive subspace signal detection in the presence of nonzero-mean clutter. By extending the traditional rank-one signal model to a more general subspace signal model, we have developed three effective detectors—SGLRT-NMC, SRao-NMC, and SAMF-NMC—based on the GLRT, Rao test, and Wald test, respectively. Theoretical analysis reveals two fundamental performance distinctions from the zero-mean clutter scenario: a reduction in the DOF of the test statistic’s distribution due to the joint estimation of the clutter mean and covariance matrix, and a SCR loss by a factor of $L/(L+1)$ resulting from estimation bias.

Extensive Monte Carlo simulations and experiments with real-world datasets have validated the analytical expressions for PD and PFA. The results demonstrate that the proposed detectors effectively suppress the adverse effects of nonzero-mean clutter, maintaining robust detection performance regardless of clutter mean power or its angular relationship with the signal. In contrast, conventional detectors SGLRT, SAMF, and SRao suffer significant performance degradation under the same conditions.
In practical applications, we recommend the SGLRT-NMC for scenarios with no signal mismatch, the SAMF-NMC for robust detection under mismatched conditions, and the SRao-NMC when high selectivity against mismatched signals is desired. No single detector universally outperforms the others across all scenarios; the choice should be guided by the operational environment and prior knowledge of signal characteristics.

Future work may explore several promising directions: 1) extension to non-Gaussian clutter models to enhance applicability in real-world environments; 2) reduction of training data requirements by exploiting historical data or structural properties of the clutter covariance matrix (e.g., persymmetry or Toeplitz structure); 3) investigation of robust designs against other types of mismatches, such as nonzero-mean clutter mismatch or time-varying clutter statistics.

\appendices 
\section{Detailed Mathematical Derivations of the Proposed Detectors}
\subsubsection{Derivations of the SGLRT-NMC}
\label{app:DrvtGLRTNMC}
We proceed to solve the denominator and numerator of \eqref{eq2} sequentially. Using Wirtinger calculus \cite{Haykin14Book} and setting the derivative of \eqref{eq3} with respect to (w.r.t.) $\mathbf{R}$ to zero yields the maximum likelihood estimate (MLE) of $\mathbf{R}$ under hypothesis ${{\text{H}}_{0}}$ for a given $\bm{\mu}$ as
\begin{equation}
	\label{eq5}
	{{\mathbf{\hat{R}}}_{0}}=\frac{1}{L+1}\bigg[ \left( \mathbf{x}-\bm{\mu}  \right){{\left( \mathbf{x}-\bm{\mu}  \right)}^{\text{H}}}+\sum\limits_{{\ell}=1}^{L}{\left( {{\mathbf{x}}_{\ell}}-\bm{\mu}  \right){{\left( {{\mathbf{x}}_{\ell}}-\bm{\mu}  \right)}^{\text{H}}}} \bigg].
\end{equation}
Substituting \eqref{eq5} into \eqref{eq3}, we obtain
\begin{equation}
	\label{eq6}
	\begin{aligned}
	{{f}_{0}}&(\mathbf{x},{{\mathbf{X}}_{L}};{{\mathbf{\hat{R}}}_{0}})=
{{{{\left( L+1 \right)}^{L+1}}}}\big/{{{{\left( \text{e}\pi  \right)}^{N(L+1)}}}\;}\\
&\cdot {{{\Big| \left( \mathbf{x}-\bm{\mu}  \right){{\left( \mathbf{x}-\bm{\mu}  \right)}^{\text{H}}}}}} {{{+\sum\limits_{{\ell}=1}^{L}{\left( {{\mathbf{x}}_{\ell}}-\bm{\mu}  \right){{\left( {{\mathbf{x}}_{\ell}}-\bm{\mu}  \right)}^{\text{H}}}} \Big|}^{-(L+1)}}}.
	\end{aligned}
\end{equation}
Nulling the partial derivative of \eqref{eq6} w.r.t. $\bm{\mu}$, we obtain the MLE of $\bm{\mu}$ under the hypothesis ${{\text{H}}_{0}}$, as shown in \eqref{eq7}.
Substituting \eqref{eq7} into \eqref{eq5} produces
\begin{equation}
	\label{eq8}
	{{\mathbf{\hat{R}}}_{0}}=\frac{1}{L+1}\left[ \left( \mathbf{x}-{{{\hat{\bm{\mu} }}}_{0}} \right){{\left( \mathbf{x}-{{{\hat{\bm{\mu} }}}_{0}} \right)}^{\text{H}}}+{{\mathbf{S}}_{0}} \right],
\end{equation}
where ${{\mathbf{S}}_{0}}$ is given in \eqref{eq9}.
Substituting \eqref{eq7} into \eqref{eq6} and applying the matrix determinant lemma, we obtain
\begin{equation}
	\label{eq10}
	\begin{aligned}
	{{f}_{0}}&(\mathbf{x},{{\mathbf{X}}_{L}};{{\mathbf{\hat{R}}}_{0}},{{\hat{\bm{\mu} }}_{0}})=\left.{{{\left( L+1 \right)}^{L+1}}}\right/{{{\left( \text{e}\pi  \right)}^{N(L+1)}}}\\
&\cdot{{{\left| {{\mathbf{S}}_{0}} \right|}^{-(L+1)}}{{\left[ 1+{{\left( \mathbf{x}-{{{\hat{\bm{\mu} }}}_{0}} \right)}^{\text{H}}}\mathbf{S}_{0}^{-1}\left( \mathbf{x}-{{{\hat{\bm{\mu} }}}_{0}} \right) \right]}^{-(L+1)}}}.
	\end{aligned}
\end{equation}

Next, we solve the numerator of \eqref{eq2}. To obtain the MLE of $\mathbf{R}$ under ${{\text{H}}_{1}}$ for given $\mathbf{\bm{\alpha}}$ and $\bm{\mu}$, we take the derivative of \eqref{eq4} w.r.t. $\mathbf{R}$ to zero, yielding
\begin{equation}
	\label{eq11}
	\begin{aligned}
	{{\mathbf{\hat{R}}}_{\bm{\alpha},\bm{\mu},1}}=\frac{1}{L+1}\bigg[ {{\mathbf{x}}_{\text{rs}}}{{\mathbf{x}}_{\text{rs}} ^{\text{H}}}+\sum\limits_{{\ell}=1}^{L}{\left( {{\mathbf{x}}_{\ell}}-\bm{\mu}  \right){{\left( {{\mathbf{x}}_{\ell}}-\bm{\mu}  \right)}^{\text{H}}}} \bigg].
	\end{aligned}
\end{equation}
Substituting \eqref{eq11} into \eqref{eq4}, we obtain
\begin{equation}
	\label{eq12}
	\begin{aligned}
	{{f}_{1}}&(\mathbf{x},{{\mathbf{X}}_{L}};{{\mathbf{\hat{R}}}_{\bm{\alpha},\bm{\mu},1}})= \left.{{{{\left( L+1 \right)}^{L+1}}}}\right/ {{{{\left( \text{e}\pi  \right)}^{N(L+1)}}}\;}\\
&\cdot{{{\Bigg|  {{\mathbf{x}}_{\text{rs}}}{{\mathbf{x}}_{\text{rs}}^{\text{H}}}}}}+{{{\sum\limits_{{\ell}=1}^{L}{\left( {{\mathbf{x}}_{\ell}}-\bm{\mu}  \right){{\left( {{\mathbf{x}}_{\ell}}-\bm{\mu}  \right)}^{\text{H}}}} \Bigg|}^{-(L+1)}}}.
	\end{aligned}
\end{equation}
Nulling the partial derivative of \eqref{eq12} w.r.t. $\bm{\mu}$, we obtain the MLE of $\bm{\mu}$ under hypothesis ${{\text{H}}_{1}}$ given $\mathbf{\bm{\alpha}}$ as 
\begin{equation}
	\label{eq13}
	{{\hat{\bm{\mu} }}_{1}}=\frac{1}{L+1}\left( \mathbf{x}-\mathbf{A\bm{\alpha} }+\sum\limits_{{\ell}=1}^{L}{{{\mathbf{x}}_{\ell}}} \right)={{\hat{\bm{\mu} }}_{0}}-\frac{1}{L+1}\mathbf{A\bm{\alpha} }.
\end{equation}
Substituting \eqref{eq13} directly into \eqref{eq12} and setting its derivative w.r.t. $\mathbf{\bm{\alpha}}$ to zero would, in principle, yield the MLE of $\mathbf{\bm{\alpha}}$. However, due to the complexity involving multiple summation terms, it is challenging to acquire an analytical solution for the MLE of $\mathbf{\bm{\alpha}}$. To address this, we first reformulate ${{\mathbf{\hat{R}}}_{\bm{\alpha},\bm{\mu},1}}$ in \eqref{eq11}.
Substituting \eqref{eq13} into \eqref{eq11}, we obtain
\begin{equation}
	\label{eq14}
	{{\mathbf{\hat{R}}}_{\bm{\alpha},1}}=\frac{1}{L+1}\left[ \left( \mathbf{x}-\mathbf{A\bm{\alpha} }-{{{\hat{\bm{\mu} }}}_{1}} \right){{\left( \mathbf{x}-\mathbf{A\bm{\alpha} }-{{{\hat{\bm{\mu} }}}_{1}} \right)}^{\text{H}}}+{{\mathbf{S}}_{1}} \right],
\end{equation}
where
\begin{equation}
	\label{eq15}
	{{\mathbf{S}}_{1}}=\sum\limits_{{\ell}=1}^{L}{({{\mathbf{x}}_{\ell}}-{{{\hat{\bm{\mu} }}}_{1}}){{({{\mathbf{x}}_{\ell}}-{{{\hat{\bm{\mu} }}}_{1}})}^{\text{H}}}}
\end{equation}
can be taken as the SCM with the nonzero mean ${\bm{\mu} }$ being removed under hypothesis $\text{H}_1$.
Substituting \eqref{eq13} into \eqref{eq14}, we have
\begin{equation}
	\begin{aligned}
{{{\mathbf{\hat{R}}}}_{\bm{\alpha},1}}&
=\frac{1}{L+1}\bigg[ ( \mathbf{x}-{{{\hat{\bm{\mu} }}}_{0}} ){{(\mathbf{x}-{{{\hat{\bm{\mu} }}}_{0}} )}^{\text{H}}}-\frac{L}{L+1}(\mathbf{x}-{{{\hat{\bm{\mu} }}}_{0}}){{\mathbf{\bm{\alpha} }}^{\text{H}}}{{\mathbf{A}}^{\text{H}}}\\
&-\frac{L}{L+1}\mathbf{A\bm{\alpha} }{{( \mathbf{x}-{{{\hat{\bm{\mu} }}}_{0}} )}^{\text{H}}}+\frac{{{L}^{2}}}{{{(L+1)}^{2}}}\mathbf{A\bm{\alpha}}{{\mathbf{\bm{\alpha} }}^{\text{H}}}{{\mathbf{A}}^{\text{H}}}+{{\mathbf{S}}_{1}} \bigg] .
	\end{aligned}
	\label{eq17}
\end{equation}
To facilitate the derivation w.r.t. $\mathbf{\bm{\alpha}}$, we first reorganize ${{\mathbf{S}}_{1}}$ in \eqref{eq17}. Substituting \eqref{eq13} into \eqref{eq15}, we obtain
\begin{equation}
	\label{eq18}
	\begin{aligned}\mathbf{S}_{1}&=\mathbf{S}_0+\frac{1}{L+1}\left(\sum\nolimits_{{\ell}=1} ^L\mathbf{x}_{\ell}-L\hat{\mathbf{\bm{\mu}}}_0\right) \mathbf{\bm{\alpha}}^\mathrm{H}\mathbf{A}^\mathrm{H}\\
&+\frac{1}{L+1}\mathbf{A}\mathbf{\bm{\alpha}}\left(\sum\nolimits_{{\ell}=1}^L \mathbf{x}_{\ell}-L\hat{\mathbf{\bm{\mu}}}_0\right)^\mathrm{H}+ \frac{L}{\left(L+1\right)^2}\mathbf{A}\mathbf{\bm{\alpha}} \mathbf{\bm{\alpha}}^\mathrm{H}\mathbf{A}^\mathrm{H}.\end{aligned}
\end{equation}
From \eqref{eq7}, we observe that 
\begin{equation}
	\label{eq19}
	\sum\limits_{{\ell}=1}^{L}{{{\mathbf{x}}_{\ell}}-L{{{\hat{\bm{\mu} }}}_{0}}}={{\hat{\bm{\mu} }}_{0}}-\mathbf{x}.
\end{equation}
Substituting \eqref{eq19} into \eqref{eq18}, we obtain
\begin{equation}
	\label{eq20}\\
	\begin{aligned}
	{{\mathbf{S}}_{1}}&={{\mathbf{S}}_{0}}-\frac{1}{L+1}\left( \mathbf{x}-{{{\hat{\bm{\mu} }}}_{0}} \right){{\mathbf{\bm{\alpha} }}^{\text{H}}}{{\mathbf{A}}^{\text{H}}}\\&-\frac{1}{L+1}\mathbf{A\bm{\alpha} }{{\left( \mathbf{x}-{{{\hat{\bm{\mu} }}}_{0}} \right)}^{\text{H}}}+\frac{L}{{{\left( L+1 \right)}^{2}}}\mathbf{A\bm{\alpha} }{{\mathbf{\bm{\alpha} }}^{\text{H}}}{{\mathbf{A}}^{\text{H}}}.
	\end{aligned}
\end{equation}
Substituting \eqref{eq20} into \eqref{eq17}, we obtain
\begin{equation}
	\label{eq21}
	\begin{aligned}\hat{\mathbf{R}}_{\bm{\alpha},1}
=\frac{1}{\left(L+1\right)} \left(\mathbf{S}_2+\mathbf{u}\mathbf{v}^\mathrm{H}\right),
\end{aligned}
\end{equation}
where
${{\mathbf{S}}_{2}}$ is given in \eqref{eq22},
and $\mathbf{u} = \frac{L+1}{L} (\mathbf{x} - \hat{\bm{\mu}}_{0}) - \mathbf{A}\bm{\alpha}$, $\mathbf{v} = (\mathbf{x} - \hat{\bm{\mu}}_{0}) - \frac{L}{L+1} \mathbf{A}\bm{\alpha}.$
Taking the determinant of the matrix in \eqref{eq21} yields
\begin{equation}
	\label{eq23}
	\begin{aligned}|\hat{\mathbf{R}}_{\bm{\alpha},1}| =\frac{1}{L+1}|\mathbf{S}_2|\left(1+\mathbf{v}^\mathrm{H}\mathbf{S}_2^{-1}\mathbf{u}\right).
\end{aligned}
\end{equation}
According to \eqref{eq11} and \eqref{eq23}, we can rewrite \eqref{eq12} as
\begin{equation}
	\label{eq24}
	\begin{aligned}
	&{{f}_{1}}(\mathbf{x},{{\mathbf{X}}_{L}};{{\mathbf{\hat{R}}}_{\bm{\alpha},1}},{{\hat{\bm{\mu} }}_{1}})={{{{\left( L+1 \right)}^{L+1}}}}\\
&\cdot \left[{{{\left( \text{e}\pi  \right)}^{N(L+1)}}{{\left| {{\mathbf{S}}_{2}} \right|}^{L+1}}\left(1+\mathbf{v}^\mathrm{H}\mathbf{S}_2^{-1}\mathbf{u}\right)^{L+1}}\right]^{-1}.
	\end{aligned}
\end{equation}
Setting the partial derivative of \eqref{eq24} w.r.t. $\mathbf{\bm{\alpha}}$ equal to zero yields the MLE of $\mathbf{\bm{\alpha}}$ as
\begin{equation}
	\label{eq25}
	\mathbf{\hat{\bm{\alpha}}}=\frac{L+1}{L}{{({{\mathbf{A}}^{\text{H}}}\mathbf{S}_{2}^{-1} \mathbf{A})} ^{-1}} {{\mathbf{A}}^{\text{H}}}\mathbf{S}_{2}^{-1}\left( \mathbf{x}-{{{\hat{\bm{\mu} }}}_{0}} \right).
\end{equation}
Substituting \eqref{eq25} into \eqref{eq24}, we obtain
\begin{equation}
	\label{eq26}
	\begin{aligned}
		&f_1(\mathbf{x},\mathbf{X}_L;\hat{\mathbf{R}}_1,\hat{\bm{\mu}}_1,\hat{\bm{\alpha}})\\ &=\frac{(L+1)^{L+1}}{(\mathrm{e}\pi)^{N(L+1)}|\mathbf{S}_2|^{L+1}}\Bigg\{1+\frac{L+1}{L}\Big[(\mathbf{x}-\hat{\bm{\mu}}_0)^{\mathrm{H}}\mathbf{S}_2^{-1}(\mathbf{x}-\hat{\bm{\mu}}_0)\\&-(\mathbf{x}-\hat{\bm{\mu}}_0)^\mathrm{H}\mathbf{S}_2^{-1}\mathbf{A}(\mathbf{A}^\mathrm{H}\mathbf{S}_2^{-1}\mathbf{A})^{-1}\mathbf{A}^\mathrm{H}\mathbf{S}_2^{-1}(\mathbf{x}-\hat{\bm{\mu}}_0)\Big]\Bigg\}.
	\end{aligned}
\end{equation}
Taking the $(L+1)$-th root of the ratio between \eqref{eq26} and \eqref{eq10} leads to the final GLRT shown in \eqref{eq27}.

\subsubsection{Derivations of the SRao-NMC}
\label{app:DrvtRaoNMC}
According to the definition in \eqref{eq30}, the FIM in \eqref{eq29} can be partitioned as
\begin{equation}
	\mathbf{I}(\mathbf{\Theta})=\begin{bmatrix} \mathbf{I}_{\bm{\alpha},\bm{\alpha}}(\mathbf{\Theta})& \mathbf{I}_{\bm{\alpha},\bm{\mu}}(\mathbf{\Theta})& \mathbf{I}_{\bm{\alpha},\mathbf{R}} (\mathbf{\Theta})\\\mathbf{I}_{\bm{\mu},\bm{\alpha}}(\mathbf{\Theta})&\mathbf{I}_{\bm{\mu}, \bm{\mu}}(\mathbf{\Theta})&\mathbf{I}_{\bm{\mu},\mathbf{R}}(\mathbf{\Theta})\\
\mathbf{I}_{\mathbf{R},\bm{\alpha}}(\mathbf{\Theta})& \mathbf{I}_{\mathbf{R},\bm{\mu}} (\mathbf{\Theta})&\mathbf{I}_{\mathbf{R},\mathbf{R}}(\mathbf{\Theta}).\end{bmatrix}
	\label{eq31}
\end{equation}

From \eqref{eq4}, we can obtain the following two equations
\begin{equation}
	\label{eq32}
\frac{\partial \ln {{f}_{1}}\left( \mathbf{x},{{\mathbf{X}}_{L}} \right)}{\partial \mathbf{\bm{\alpha} }}={{(\mathbf{x}_{\text{rs}}^{\text{H}}{{\mathbf{R}}^{-1}}\mathbf{A})}^{\text{T}}},
\end{equation}
\begin{equation}
	\label{eq33}
\frac{\partial \ln {{f}_{1}}\left( \mathbf{x},{{\mathbf{X}}_{L}} \right)}{\partial {{\mathbf{\bm{\alpha} }}^{*}}}={{\mathbf{A}}^{\text{H}}}{{\mathbf{R}}^{-1}}{{\mathbf{x}}_{\text{rs}}}.
\end{equation}
Substituting \eqref{eq32} and \eqref{eq33} into \eqref{eq29} yields
\begin{equation}
	\label{eq34}
	\begin{aligned}
	 {{\mathbf{I}}_{\mathbf{\bm{\alpha} },\mathbf{\bm{\alpha} }}}(\mathbf{\Theta }) ={{\mathbf{A}}^{\text{H}}}{{\mathbf{R}}^{-1}}\mathbf{A}.
	\end{aligned}
\end{equation}
Taking the partial derivative of \eqref{eq32} w.r.t. $\mathbf{R}$ and computing its expectation yields ${{\mathbf{I}}_{\mathbf{\bm{\alpha}},\mathbf{R}}}(\mathbf{\Theta})$ as a zero vector. Similarly, taking the partial derivative of \eqref{eq33} w.r.t. ${{\bm{\mu}}^{\text{T}}}$ gives
\begin{equation}
	\label{eq35}
	\frac{\partial \ln {{f}_{1}}\left( \mathbf{x},{{\mathbf{X}}_{L}} \right)}{\partial {{\mathbf{\bm{\alpha} }}^{*}}\partial {{\bm{\mu} }^{\text{T}}}}=-{{\mathbf{A}}^{\text{H}}}{{\mathbf{R}}^{-1}}.
\end{equation}
From \eqref{eq4}, we can derive the following two equations
\begin{equation}
	\label{eq36}
	\frac{\partial \ln {{f}_{1}}\left( \mathbf{x},{{\mathbf{X}}_{L}} \right)}{\partial {{\bm{\mu} }^{*}}}={{\mathbf{R}}^{-1}} {{\mathbf{x}}_{\text{rs}}}+ \sum\nolimits_{{\ell}=1}^{L}{{\mathbf{R}}^{-1}} {({\mathbf{x}}_{\ell}-\bm{\mu})} ,
\end{equation}
\begin{equation}
	\label{eq37}
	\frac{\partial \ln {{f}_{1}}\left( \mathbf{x},{{\mathbf{X}}_{L}} \right)}{\partial {{\bm{\mu} }^{\text{T}}}}={{\mathbf{x}}_{\text{rs}}^{\text{H}}}{{\mathbf{R}}^{-1}}+ \sum\nolimits_{{\ell}=1}^{L}{({\mathbf{x}}_{\ell}-\bm{\mu})}^{\text{H}}{{\mathbf{R}}^{-1}} .
\end{equation}
Substituting \eqref{eq36} and \eqref{eq37} into \eqref{eq29} yields
\begin{equation}
	\label{eq38}
	{{\mathbf{I}}_{\bm{\mu} ,\bm{\mu} }}(\mathbf{\Theta })=
(L+1){{\mathbf{R}}^{-1}}.
\end{equation}
Taking the partial derivative of \eqref{eq33} w.r.t. $\text{vec}^{\text{T}}(\mathbf{R})$ and computing its expectation yields
\begin{equation}
	\label{eq39}
	{{\mathbf{I}}_{\mathbf{\bm{\alpha} },\mathbf{R}}}(\mathbf{\Theta })
={\mathbf{0}_{p\times {{N}^{2}}}}.
\end{equation}
Computing the partial derivative of \eqref{eq36} w.r.t. $\text{vec}^{\text{T}}(\mathbf{R})$ and taking its expectation yields
\begin{equation}
	\label{eq40}
	{{\mathbf{I}}_{\bm{\mu} ,\mathbf{R}}}(\mathbf{\Theta })
={\mathbf{0}_{N\times {{N}^{2}}}}.
\end{equation}
Substituting \eqref{eq34}, \eqref{eq35}, \eqref{eq38}, \eqref{eq39}, and \eqref{eq40} into \eqref{eq31} yields
\begin{equation}
	\label{eq41}
	\mathbf{I}(\mathbf{\Theta})=\begin{bmatrix}\mathbf{A}^\mathrm{H}\mathbf{R}^{-1}\mathbf{A}& -\mathbf{A}^\mathrm{H}\mathbf{R}^{-1}&\mathbf{0}_{p\times N^2}\\
-\mathbf{R}^{-1}\mathbf{A}& (L+1)\mathbf{R}^{-1}&\mathbf{0}_{N\times N^2}\\
\mathbf{0}_{N^2\times p}&\mathbf{0}_{N^2\times N}&\mathbf{I}_{\mathbf{R},\mathbf{R}}(\mathbf{\Theta})\end{bmatrix}.
\end{equation}
From \eqref{eq41}, we observe that $\mathbf{I}(\mathbf{\Theta})$ is a block diagonal matrix. Consequently, ${{\mathbf{I}}_{\mathbf{R},\mathbf{R}}}(\mathbf{\Theta})$ has no influence on ${{\left[ {{\mathbf{I}}^{-1}}(\mathbf{\Theta})\right]}_{{{\mathbf{\Theta }}_{\text{r}}}{{\mathbf{\Theta }}_{\text{r}}}}} $. Based on \eqref{eq41}, applying the block matrix inversion lemma and incorporating the definition in \eqref{eq31}, we obtain
\begin{equation}
	\label{eq42}
	{{\left\{ {{\left[ {{\mathbf{I}}^{-1}}(\mathbf{\Theta }) \right]}_{{{\mathbf{\Theta }}_{\text{r}}},{{\mathbf{\Theta }}_{\text{r}}}}} \right\}}^{-1}}=\frac{L}{L+1}{{\mathbf{A}}^{\text{H}}}{{\mathbf{R}}^{-1}}\mathbf{A} .
\end{equation}

By substituting \eqref{eq32}, \eqref{eq33}, and \eqref{eq42} into \eqref{eq28}, setting $\mathbf{\bm{\alpha}} = \mathbf{0}$ and ignoring the constant, we obtain the Rao test statistic for given $\bm{\mu}$ and $\mathbf{R}$
\begin{equation}
	\label{eq44}
	{{t}_{\text{Ra}{{\text{o}}_{\bm{\mu} ,\mathbf{R}}}}}={{\left( \mathbf{x}-\bm{\mu}  \right)}^{\text{H}}}{{\mathbf{R}}^{-1}}\mathbf{A}{{(\mathbf{A}_{{}}^{\text{H}}{{\mathbf{R}}^{-1}} \mathbf{A})}^{-1}}\mathbf{A}_{{}}^{\text{H}}{{\mathbf{R}}^{-1}}\left( \mathbf{x}-\bm{\mu}  \right).
\end{equation}
Substituting \eqref{eq8} into \eqref{eq44} yields the final Rao test statistic shown in \eqref{eq45}.

\section{Equivalent Forms of the Proposed Detectors}
\subsubsection{Equivalent Forms of the SGLRT-NMC}
\label{app:GLRTNMC}
Applying the matrix determinant lemma to the numerator of \eqref{eq27}, we obtain
\begin{equation}
	\label{eq50}
	\begin{aligned}
	&{{t}_{\text{SGLRT-NMC}}}=\left.{\left| {{\mathbf{S}}_{0}}+\left( \mathbf{x}-{{{\hat{\bm{\mu} }}}_{0}} \right){{\left( \mathbf{x}-{{{\hat{\bm{\mu} }}}_{0}} \right)}^{\text{H}}} \right|} \right/ {\left| {{\mathbf{S}}_{2}} \right|} \\ &\cdot\left\{ 1+\frac{L+1}{L}\left[ {{\left( \mathbf{x}-{{{\hat{\bm{\mu} }}}_{0}} \right)}^{\text{H}}}\mathbf{S}_{2}^{-1}\left( \mathbf{x}-{{{\hat{\bm{\mu} }}}_{0}} \right)\right.\right.\\&\left.\left.-{{\left( \mathbf{x}-{{{\hat{\bm{\mu} }}}_{0}} \right)}^{\text{H}}}\mathbf{S}_{2}^{-1}\mathbf{A}{{({{\mathbf{A}}^{\text{H}}}\mathbf{S}_{2}^{-1}\mathbf{A})}^{-1}}{{\mathbf{A}}^{\text{H}}}\mathbf{S}_{2}^{-1}\left( \mathbf{x}-{{{\hat{\bm{\mu} }}}_{0}} \right) \right] \right\}^{-1}.
	\end{aligned}
\end{equation}
To facilitate the analysis of the detector's statistical performance, we express $ {{\mathbf{S}}_{0}} $  in \eqref{eq50} in terms related to $ {{\mathbf{S}}_{2}} $.  Based on \eqref{eq22}, we derive
\begin{equation}
	\label{eq51}
	{{\mathbf{S}}_{0}}={{\mathbf{S}}_{2}}+\frac{1}{L}\left( \mathbf{x}-{{{\hat{\bm{\mu} }}}_{0}} \right){{\left( \mathbf{x}-{{{\hat{\bm{\mu} }}}_{0}} \right)}^{\text{H}}}.
\end{equation}
Substituting \eqref{eq51} into \eqref{eq50} and applying the matrix determinant lemma to the numerator yields
\begin{equation}
	\label{eq52}
	\begin{aligned}
	&{{t}_{\text{SGLRT-NMC}}}={\left[ 1+\frac{L+1}{L}{{\left( \mathbf{x}-{{{\hat{\bm{\mu} }}}_{0}} \right)}^{\text{H}}}\mathbf{S}_{2}^{-1}\left( \mathbf{x}-{{{\hat{\bm{\mu} }}}_{0}} \right) \right]}\\&\cdot{\left\{ 1+\frac{L+1}{L}\left[ {{\left( \mathbf{x}-{{{\hat{\bm{\mu} }}}_{0}} \right)}^{\text{H}}}\mathbf{S}_{2}^{-1}\left( \mathbf{x}-{{{\hat{\bm{\mu} }}}_{0}} \right)\right.\right.}\\&{\left.\left.-{{\left( \mathbf{x}-{{{\hat{\bm{\mu} }}}_{0}} \right)}^{\text{H}}}\mathbf{S}_{2}^{-1}\mathbf{A}{{({{\mathbf{A}}^{\text{H}}}\mathbf{S}_{2}^{-1}\mathbf{A})}^{-1}}{{\mathbf{A}}^{\text{H}}}\mathbf{S}_{2}^{-1}\left( \mathbf{x}-{{{\hat{\bm{\mu} }}}_{0}} \right) \right] \right\}^{-1}}.
	\end{aligned}
\end{equation}
The detector in \eqref{eq52} is equivalent to
\begin{equation}
	\label{eq53}
	\begin{aligned}
	&t_{\text{SGLRT-NMC}}^{'}\\&={\left[\frac{L+1}{L}{{\left( \mathbf{x}-{{{\hat{\bm{\mu} }}}_{0}} \right)}^{\text{H}}}\mathbf{S}_{2}^{-1}\mathbf{A}{{({{\mathbf{A}}^{\text{H}}}\mathbf{S}_{2}^{-1}\mathbf{A})}^{-1}}{{\mathbf{A}}^{\text{H}}}\mathbf{S}_{2}^{-1}\left( \mathbf{x}-{{{\hat{\bm{\mu} }}}_{0}} \right)\right]}\\&\cdot\left\{1+\frac{L+1}{L}\left[ {{\left( \mathbf{x}-{{{\hat{\bm{\mu} }}}_{0}} \right)}^{\text{H}}}\mathbf{S}_{2}^{-1}\left( \mathbf{x}-{{{\hat{\bm{\mu} }}}_{0}} \right)-\right.\right.\\&\left.\left.\cdot{{\left( \mathbf{x}-{{{\hat{\bm{\mu} }}}_{0}} \right)}^{\text{H}}}\mathbf{S}_{2}^{-1}\mathbf{A}{{({{\mathbf{A}}^{\text{H}}}\mathbf{S}_{2}^{-1}\mathbf{A})}^{-1}}{{\mathbf{A}}^{\text{H}}}\mathbf{S}_{2}^{-1}\left( \mathbf{x}-{{{\hat{\bm{\mu} }}}_{0}} \right) \right]\right\}^{-1},
	\end{aligned}
\end{equation}
due to $t_{\text{SGLRT-NMC}}^{'}={{t}_{\text{SGLRT-NMC}}}-1$. Using \eqref{eq54}, we can rewrite \eqref{eq53} as \eqref{eq55}.

\subsubsection{Equivalent Forms of the SRao-NMC}
\label{app:RaoNMC}
Substituting \eqref{eq51} into \eqref{eq45} and utilizing \eqref{eq54}, while neglecting the constant term, we obtain
\begin{equation}
	\label{eq56}
	\begin{aligned}
	t_{\text{SRao-NMC}}&={{\mathbf{z}}^{\text{H}}}{{(\mathbf{z}{{\mathbf{z}}^{\text{H}}}+ {{\mathbf{S}}_{2}})}^{-1}}\mathbf{A}{{\left\{ \mathbf{A}_{{}}^{\text{H}}{{(\mathbf{z}{{\mathbf{z}}^{\text{H}}}+ {{\mathbf{S}}_{2}})}^{-1}}\mathbf{A} \right\}}^{-1}}\\&\cdot\mathbf{A}_{{}}^{\text{H}} {{(\mathbf{z}{{\mathbf{z}} ^{\text{H}}}+{{\mathbf{S}}_{2}})}^{-1}}\mathbf{z}.
	\end{aligned}
\end{equation}
Taking the inverse of $ (\mathbf{z}{{\mathbf{z}}^{\text{H}}}+{{\mathbf{S}}_{2}}) $ in \eqref{eq56} yields
\begin{equation}
	\label{eq57}
	{{(\mathbf{z}{{\mathbf{z}}^{\text{H}}}+{{\mathbf{S}}_{2}})}^{-1}}= \mathbf{S}_{2}^{-1}-\frac{\mathbf{S}_{2}^{-1}\mathbf{z}{{\mathbf{z}}^{\text{H}}} \mathbf{S}_{2}^{-1}}{1+{{\mathbf{z}}^{\text{H}}}\mathbf{S}_{2}^{-1}\mathbf{z}}.
\end{equation}
Using \eqref{eq57} and the matrix inversion lemma \cite{Hayes96Book}, we can further obtain
\begin{equation}
	\label{eq58}
	\begin{aligned}
&\left\{\mathbf{A}^\mathrm{H}(\mathbf{z}\mathbf{z}^\mathrm{H}+\mathbf{S}_2)^{-1} \mathbf{A}\right\}^{-1}=(\mathbf{A}^\mathrm{H}\mathbf{S}_2^{-1}\mathbf{A})^{-1}+\\
&\frac{(\mathbf{A}^\mathrm{H}\mathbf{S}_2^{-1}\mathbf{A})^{-1}\mathbf{A}^\mathrm{H} \mathbf{S}_2^{-1}\mathbf{z}\mathbf{z}^\mathrm{H}\mathbf{S}_2^{-1}\mathbf{A}(\mathbf{A}^\mathrm{H} \mathbf{S}_2^{-1}\mathbf{A})^{-1}}{1+\mathbf{z}^\mathrm{H}\mathbf{S}_2^{-1}\mathbf{z}- \mathbf{z}^\mathrm{H}\mathbf{S}_2^{-1}\mathbf{A}(\mathbf{A}^\mathrm{H}\mathbf{S}_2^{-1} \mathbf{A})^{-1}\mathbf{A}^\mathrm{H}\mathbf{S}_2^{-1}\mathbf{z}} .
	\end{aligned}
\end{equation}
According to \eqref{eq57}, we obtain
\begin{equation}
	\label{eq59}
	\begin{aligned}
	\mathbf{A}_{{}}^{\text{H}}{{(\mathbf{z}{{\mathbf{z}}^{\text{H}}}+ {{\mathbf{S}}_{2}})}^{-1}}\mathbf{z}=\frac{\mathbf{A}_{{}}^{\text{H}} \mathbf{S}_{2}^{-1}\mathbf{z}}{1+{{\mathbf{z}}^{\text{H}}}\mathbf{S}_{2}^{-1}\mathbf{z}}.
	\end{aligned}
\end{equation}
Substituting \eqref{eq58} and \eqref{eq59} into \eqref{eq56}, after some algebra, lead to the equivalent of the SRao-NMC given in \eqref{eq60}.

\subsubsection{Equivalent Forms of the SAMF-NMC}
\label{app:WaldNMC}
Taking the inverse of \eqref{eq21} gives
\begin{equation}
	\label{eq61}
	\mathbf{\hat{R}}_{1}^{-1} = (L+1) \Bigg\{
	\mathbf{S}_{2}^{-1} - \frac{
	\mathbf{S}_{2}^{-1} \hat{\mathbf{u}} \hat{\mathbf{v}}^{\text{H}} \mathbf{S}_{2}^{-1}}{1 + \hat{\mathbf{v}}^{\text{H}} \mathbf{S}_{2}^{-1} \hat{\mathbf{u}}} \Bigg\},
\end{equation}
where,
$\hat{\mathbf{u}} = \frac{L+1}{L} (\mathbf{x} - \hat{\bm{\mu}}_{0}) - \mathbf{A}\hat{\bm{\alpha}}$, $
\hat{\mathbf{v}} = (\mathbf{x} - \hat{\bm{\mu}}_{0}) - \frac{L}{L+1} \mathbf{A}\hat{\bm{\alpha}}.$
Multiplying \eqref{eq61} on the left by ${{\mathbf{A}}^{\text{H}}}$ and on the right by $\mathbf{A}$ yields
\begin{equation}
	\label{eq62}
	{{\mathbf{A}}^{\text{H}}}\mathbf{\hat{R}}_{1}^{-1}\mathbf{A}=\left( L+1 \right)\left\{ {{\mathbf{A}}^{\text{H}}}\mathbf{S}_{2}^{-1}\mathbf{A}- \frac{{\mathbf{A}}^{\text{H}}\mathbf{S}_{2}^{-1} \hat{\mathbf{u}} \hat{\mathbf{v}}^{\text{H}} \mathbf{S}_{2}^{-1}{\mathbf{A}}}{1 + \hat{\mathbf{v}}^{\text{H}} \mathbf{S}_{2}^{-1} \hat{\mathbf{u}}}\right\}.
\end{equation}
Substituting \eqref{eq25} into \eqref{eq62} yields
\begin{equation}
	\label{eq63}
	{{\mathbf{A}}^{\text{H}}}\mathbf{\hat{R}}_{1}^{-1}\mathbf{A}=\left( L+1 \right){{\mathbf{A}}^{\text{H}}}\mathbf{S}_{2}^{-1}\mathbf{A}.
\end{equation}
By substituting \eqref{eq63} into \eqref{eq49} and omitting the constant terms, we obtain
\begin{equation}
	\label{eq64}
	{{t}_{\text{SAMF-NMC}}}={{\left( \mathbf{x}-{{{\hat{\bm{\mu} }}}_{0}} \right)}^{\text{H}}}\mathbf{S}_{2}^{-1}\mathbf{A}({{\mathbf{A}}^{\text{H}}}\mathbf{S}_{2}^{-1}\mathbf{A})_{{}}^{-1}{{\mathbf{A}}^{\text{H}}}\mathbf{S}_{2}^{-1}\left( \mathbf{x}-{{{\hat{\bm{\mu} }}}_{0}} \right).
\end{equation}
Substituting \eqref{eq54} into \eqref{eq64} and neglecting the constant terms, we arrive at \eqref{eq65}.

\section{Conditions for ${{\mathbf{S}}_{2}}$ Positive Definition}
\label{app:S2}
Substituting \eqref{eq7} into \eqref{eq9} yields
\begin{equation}
	\label{eq120}
	\begin{aligned}
	{{\mathbf{S}}_{0}}&=\sum\limits_{{\ell}=1}^{L}{\left[ {{\mathbf{x}}_{\ell}}\mathbf{x}_{\ell}^{\text{H}}+\frac{1}{{{\left( L+1 \right)}^{2}}}\mathbf{x}{{\mathbf{x}}^{\text{H}}}-\frac{1}{{{\left( L+1 \right)}^{2}}}\mathbf{xx}_{\ell}^{\text{H}}\right.}\\&{\left.-\frac{1}{{{\left( L+1 \right)}^{2}}}{{\mathbf{x}}_{\ell}}{{\mathbf{x}}^{\text{H}}} \right]}-\frac{L+2}{{{\left( L+1 \right)}^{2}}}\sum\limits_{{\ell}=1}^{L}{{{\mathbf{x}}_{\ell}}}\sum\limits_{{\ell}=1}^{L} {\mathbf{x}_{\ell}^{\text{H}}}.
	\end{aligned}
\end{equation}
Substituting \eqref{eq7} into \eqref{eq22} leads to
\begin{equation}
	\label{eq121}\\
	\begin{aligned}
	{{\mathbf{S}}_{2}}&={{\mathbf{S}}_{0}}-\frac{1}{L}\left[ \frac{{{L}^{2}}}{{{\left( L+1 \right)}^{2}}}\mathbf{x}{{\mathbf{x}}^{\text{H}}}-\frac{L}{{{\left( L+1 \right)}^{2}}}\mathbf{x}\sum\limits_{{\ell}=1}^{L}{\mathbf{x}_{\ell}^{\text{H}}}\right.\\ &\left.-\frac{L}{{{\left( L+1 \right)}^{2}}}\sum\limits_{{\ell}=1}^{L}{{{\mathbf{x}}_{\ell}}{{\mathbf{x}}^{\text{H}}}}+ \frac{1}{{{\left( L+1 \right)}^{2}}}\sum\limits_{{\ell}=1}^{L}{{{\mathbf{x}}_{\ell}}} \sum\limits_{{\ell}=1}^{L}{\mathbf{x}_{\ell}^{\text{H}}} \right].
	\end{aligned}
\end{equation}
Substituting \eqref{eq120} into  \eqref{eq121} results in
\begin{equation}
	\label{eq122}
	\begin{aligned}\mathbf{S}_{2}=\sum_{{\ell}=1}^{L}\mathbf{x}_{\ell}\mathbf{x}_{\ell}^{\mathrm{H}} -\frac{1}{L}\sum_{{\ell}=1}^{L}\mathbf{x}_{\ell}\sum_{{\ell}=1}^{L}\mathbf{x}_{\ell}^{\mathrm{H}}
=\begin{pmatrix}L-1\end{pmatrix}\hat{\mathbf{R}},\end{aligned}
\end{equation}
where $\mathbf{\hat{R}}=\frac{1}{L-1}\sum\limits_{{\ell}=1}^{L}{\left( {{\mathbf{x}}_{\ell}}-\frac{1}{L}\sum\nolimits_{i=1}^{L}{{{\mathbf{x}}_{i}}} \right){{\left( {{\mathbf{x}}_{\ell}}-\frac{1}{L}\sum\nolimits_{i=1}^{L}{{{\mathbf{x}}_{i}}} \right)}^{\text{H}}}}$
denotes the SCM\footnote{In radar signal processing, ${{\mathbf{\hat{R}}}^{'}}=\frac{1}{L}\sum\limits_{{\ell}=1}^{L}{{{\mathbf{x}}_{\ell}}\mathbf{x}_{\ell} ^{\text{H}}}$ or ${{\mathbf{S}}}=\sum\limits_{{\ell}=1}^{L}{{{\mathbf{x}}_{\ell}} \mathbf{x}_{\ell}^{\text{H}}}$ is termed SCM, which is positive definite with probability 1 when $L \geq N$}. According to Theorem 3.1.4 in \cite{Muirhead05}, $\mathbf{\hat{R}}$ is positive definite with probability 1 when $L \ge N+1$.

\section{Gradient Test-Based Detector}
\label{app:GradTest}
The gradient test is \cite{SunLiu22TSP}
\begin{equation}
	\label{eq124}
	{{t}_{\text{Gradient}}}=\text{Re}\left\{ {{\frac{\partial \ln {{f}_{1}}\left( \mathbf{x},{{\mathbf{X}}_{L}} \right)}{\partial \mathbf{\Theta }_{\text{r}}^{\text{T}}} \Bigg|}_{\mathbf{\Theta }={{{\mathbf{\hat{\Theta }}}}_{0}}}}\left( {{{\mathbf{\hat{\Theta }}}}_{{{\text{r}}_{1}}}}-{{\mathbf{\Theta }}_{{{\text{r}}_{0}}}} \right) \right\}.
\end{equation}
Substituting \eqref{eq25}, \eqref{eq54}, and \eqref{eq32} into \eqref{eq124} while neglecting the constant terms yields
\begin{equation}
	\label{eq125}
	{{t}_{\text{Gradient}}}=\text{Re}\left\{ {{\mathbf{z}}^{\text{H}}}\mathbf{\hat{R}}_{0}^{-1}\mathbf{A}{{({{\mathbf{A}}^{\text{H}}} \mathbf{S}_{2}^{-1}\mathbf{A})}^{-1}}{{\mathbf{A}}^{\text{H}}}\mathbf{S}_{2}^{-1}\mathbf{z} \right\}.
\end{equation}
According to \eqref{eq8}, \eqref{eq54} and \eqref{eq22}, we can obtain
\begin{equation}
	\label{eq126}
	{{\mathbf{\hat{R}}}_{0}}=\frac{1}{L+1}\left( \mathbf{z}{{\mathbf{z}}^{\text{H}}}+{{\mathbf{S}}_{2}} \right).
\end{equation}
The inverse of \eqref{eq126} is given by \eqref{eq57}. Substituting \eqref{eq57} into \eqref{eq125} and neglecting the constant terms yields
\begin{equation}
	\label{eq127}
	{{t}_{\text{Gradient}}}=\frac{\mathbf{z}^\mathrm{H}\mathbf{S}_2^{-1}\mathbf{A} (\mathbf{A}^\mathrm{H}\mathbf{S}_2^{-1}\mathbf{A})^{-1}\mathbf{A}^\mathrm{H}\mathbf{S}_2^{-1} \mathbf{z}}{1+\mathbf{z}^\mathrm{H}\mathbf{S}_2^{-1}\mathbf{z}}.
\end{equation}
It is easy to verify that $ 0<{{t}_{\text{Gradient}}}<1 $ and
\begin{equation}
	\label{eq128}
	t_{\text{GLRT}}^{'}= {{{t}_{\text{Gradient}}}} \big/{(1-{{t}_{\text{Gradient}}})},
\end{equation}
where $ t_{\text{GLRT}}^{'} $ is given by \eqref{eq55}. Since detection statistics that are mutually monotonically increasing functions are equivalent \cite{LiuLiu22SCIS}, \eqref{eq128} demonstrates the equivalence between the gradient test statistic and the GLRT-based detector.

\section{Durbin Test-Based Detector}
\label{app:DurbinTest}
The Durbin is \cite{SunLiu22TSP}
\begin{equation}
	\label{eq129}
	\begin{aligned}
	{{t}_{\text{Durbin}}}&={{ ( {{{\mathbf{\hat{\Theta }}}}_{\text{r},01}}-{{\mathbf{\Theta }}_{\text{r0}}} )}^{\text{H}}}\left\{ {{\left[ \mathbf{I}({{{\mathbf{\hat{\Theta }}}}_{0}}) \right]}_{{{\mathbf{\Theta }}_{\text{r}}},{{\mathbf{\Theta }}_{\text{r}}}}}{{\left[ {{\mathbf{I}}^{-1}}({{{\mathbf{\hat{\Theta }}}}_{0}}) \right]}_{{{\mathbf{\Theta }}_{\text{r}}},{{\mathbf{\Theta }}_{\text{r}}}}}\right.\\&\cdot\left.{{\left[ \mathbf{I}({{{\mathbf{\hat{\Theta }}}}_{0}}) \right]}_{{{\mathbf{\Theta }}_{\text{r}}},{{\mathbf{\Theta }}_{\text{r}}}}} \right\} ( {{{\mathbf{\hat{\Theta }}}}_{\text{r},01}}-{{\mathbf{\Theta }}_{\text{r0}}}  ),
	\end{aligned}
\end{equation}
where ${{\mathbf{\hat{\Theta }}}_{\text{r},01}}\triangleq \underset{{{\mathbf{\Theta }}_{\text{r}}}}{\mathop{\text{argmax}}}\,\text{ }{{f}_{1}}\left( \mathbf{x},\mathbf{X}_L|{{\mathbf{\Theta }}_{\text{r}}},{{{\mathbf{\hat{\Theta }}}}_{\text{s0}}} \right)$,
$ {{\mathbf{\hat{\Theta }}}_{\text{s}0}} $ denotes the MLE of $ {{\mathbf{\Theta }}}_{\text{s}} $ under hypothesis ${{\text{H}}_{0}}$.

From \eqref{eq4}, when $ \bm{\mu} $ and $ \mathbf{R} $ are given, the MLE of $ \bm{\alpha} $ is
\begin{equation}
	\label{eq132}
	{{\mathbf{\hat{\bm{\alpha} }}}_{\bm{\mu} ,\mathbf{R}}}={{({{\mathbf{A}}^{\text{H}}}{{\mathbf{R}}^{-1}}\mathbf{A})}^{-1}}{{\mathbf{A}}^ {\text{H}}}{{\mathbf{R}}^{-1}}\left( \mathbf{x}-\bm{\mu}  \right).
\end{equation}
By substituting \eqref{eq42} and \eqref{eq132} into \eqref{eq129} and ignoring the constant, we obtain the Durbin test statistic under the condition that both $\bm{\mu}$ and $\mathbf{R}$ are given as
\begin{equation}
	\label{eq133}
	{{t}_{\text{Durbi}{{\text{n}}_{\bm{\mu} ,\mathbf{R}}}}}={{(\mathbf{x}-\bm{\mu} )}^{\text{H}}}{{\mathbf{R}}^{-1}}\mathbf{A}{{({{\mathbf{A}}^{\text{H}}}{{\mathbf{R}}^{-1}} \mathbf{A})}^{-1}}{{\mathbf{A}}^{\text{H}}}{{\mathbf{R}}^{-1}}(\mathbf{x}-\bm{\mu} ).
\end{equation}
Equations \eqref{eq44} and \eqref{eq133} show that the Rao test statistic and Durbin test statistic share the same structure. It follows that substituting \eqref{eq7} and \eqref{eq8} into \eqref{eq133} yields identical results to \eqref{eq45}. Therefore, the Durbin test statistic demonstrates mathematical equivalence with the Rao test statistic.

\section{Generation Methods for Signal Matrix $\mathbf{A}$, True
Signal Component $\mathbf{p}_{0}$, and Clutter's Nonzero Mean $\bm{\mu}$ }
\label{app:ParameterGeneration}
\begin{enumerate}
    \item Generate the signal matrix $\mathbf{A}$.
    Specifically, $\mathbf{A}$ has the form
    $
    \mathbf{A} = \left[ \mathbf{a}_1, \mathbf{a}_2, \ldots, \mathbf{a}_p \right],
    $
    where $\mathbf{a}_m$, $m = 1, 2, \ldots, p$, follows the array manifold structure:
    $
    \mathbf{a}_m = \left[ 1, e^{-j2\pi f_m}, e^{-j2\pi 2 f_m}, \ldots, e^{-j2\pi (N-1) f_m } \right]^{\text{T}}.
    $
    Here, $f_m$ denotes the normalized frequency, which is randomly generated within the interval $(-0.5, 0.5)$. The normalized frequencies $f_m$ are randomly generated within $(-0.5, 0.5)$ to ensure that the column vectors of the signal subspace are linearly independent, which satisfies the assumption of full-rank signal subspace.

    \item Randomly generate a $p \times 1$-dimensional column vector $\bm{\alpha}_0$.
    First, compute the intermediate vector
    $
    \mathbf{p}_{\text{tmp}'} = \mathbf{A}\mathbf{\bm\alpha}_0.
    $
    Then, normalize $\mathbf{p}_{\text{tmp}'}$ by its Euclidean norm to get:
    $
    \mathbf{p}_{\text{tmp}} = \frac{\mathbf{p}_{\text{tmp}'}}{\|\mathbf{p}_{\text{tmp}'}\|}
    $,
    where $\|\cdot\|$ represents the Euclidean norm of a vector.

    \item Perform singular value decomposition (SVD) \cite{Petersen12Book} on $\mathbf{R}^{-1}\mathbf{A}$, which gives
    $
    \mathbf{R}^{-1}\mathbf{A} = \mathbf{U}\mathbf{\Sigma}\mathbf{V}
    $.
    In the decomposition, $\mathbf{U}$ and $\mathbf{V}$ are the left and right unitary matrices, respectively, and $\mathbf{\Sigma}$ is a diagonal matrix containing the singular values of $\mathbf{R}^{-1}\mathbf{A}$.
    Let $\mathbf{p}_{\perp} = \mathbf{u}_{\text{ed}}$, where $\mathbf{u}_{\text{ed}}$ denotes the last column of $\mathbf{U}$.

    \item Generate 300 weight coefficients $w_i$, $i = 1, 2, \ldots, 300$, uniformly within the interval $[0, 1]$.
    Using these weights, construct candidate true target steering vectors as
    $
    \mathbf{p}_0(i) = w_i \mathbf{p}_{\text{tmp}} + (1 - w_i)  \mathbf{p}_{\perp}
    $.

    \item For each of the 300 candidate steering vectors $\mathbf{p}_0(i)$, calculate the corresponding $\cos^2\theta$ value using \eqref{eq74} in the main text.
    Denote the $i$-th calculated value as $\cos^2\theta(i)$.

    \item Let $\cos^2\theta_*$ be the preset target value of $\cos^2\theta$.
    Select the weight coefficient that minimizes the absolute difference $|\cos^2\theta(i) - \cos^2\theta_*|$, and denote this optimal weight as $w_{\text{opt}}$.
    The true target steering vector before amplitude scaling is then
    $
    \mathbf{p}_{0,n} = w_{\text{opt}} \mathbf{p}_{\text{tmp}} + (1 - w_{\text{opt}}) \mathbf{p}_{\perp}
    $.

    \item Compute the quantity
    $
    \rho_n = \mathbf{p}_{0,n}^{\text{H}} \mathbf{R}^{-1} \mathbf{p}_{0,n}
    $.
    Then, calculate the amplitude scaling factor for the target signal
    $
    \lambda_p = \sqrt{\tfrac{(L+1)\rho}{L\rho_n}}
    $. It follows that the true target signal that satisfies the specific SCR in \eqref{eq75} is
    $
    \mathbf{p}_0 = \lambda_p \mathbf{p}_{0,n}
    $.

    \item Generate 500 candidate normalized frequencies $f_c(k)$, $k = 1, 2, \ldots, 500$, within the interval $[-0.5, 0.5)$.
    For each $f_c(k)$, construct the corresponding candidate steering vector for nonzero-mean clutter:
    $
    \bm{\mu}_0(k) = \left[ 1, e^{-j2\pi f_c(k)}, e^{-j2\pi 2 f_c(k)}, \ldots, e^{-j2\pi (N-1) f_c(k)} \right]^T
    $.

    \item For each of the 500 candidate clutter steering vectors $\bm{\mu}_0(k)$, calculate the corresponding $\cos^2\phi$ value using \eqref{eq109}.
    Denote the $k$-th calculated value as $\cos^2\phi(k)$.

    \item Let $\cos^2\phi_*$ be the preset target value of $\cos^2\phi$.
    Select the normalized frequency $f_c(k)$ that minimizes the absolute difference $|\cos^2\phi(k) - \cos^2\phi_*|$, and denote this optimal frequency as $f_c$. It follows that the steering vector of the nonzero-mean clutter is
    $
    \bm{\mu}_0 = \left[ 1, e^{-j2\pi f_c}, e^{-j2\pi 2 f_c }, \ldots, e^{-j2\pi (N-1) f_c} \right]^T
    $.

    \item Let $\xi$ be the preset amplitude power of nonzero-mean clutter in \eqref{eq108}.
    First, compute the amplitude scaling factor for the clutter
    $
    \lambda_c = \sqrt{\frac{\xi}{\bm{\mu}_0^\text{H} \mathbf{R}^{-1} \bm{\mu}_0}}
    $. Then, the clutter's final nonzero mean is
    $
    \bm{\mu} = \lambda_c  \bm{\mu}_0
    $.
\end{enumerate}

\bibliographystyle{IEEEtran} 
\bibliography{D:/LaTexReference/Detection}
\end{document}